\pdfoutput=1
\documentclass[11pt,aps,prd,a4paper,showpacs,showkeys,superscriptaddress, preprintnumbers,floatfix,nofootinbib,notitlepage]{revtex4-1}

\usepackage[toc,page]{appendix}
\usepackage{latexsym}
\usepackage{revsymb}
\usepackage{multirow}
\usepackage{color}
\usepackage[usenames,dvipsnames,svgnames,table]{xcolor}

\usepackage{graphicx}
\usepackage{epsfig}  
\usepackage{epsf}    
\usepackage{dcolumn}
\usepackage{bm}
\usepackage{textcomp}
\usepackage{float}
\usepackage[]{hyperref}
\usepackage{adjustbox}
\usepackage[utf8]{inputenc}
\usepackage[T1]{fontenc} 
\usepackage{url}
\usepackage{amsmath}
\usepackage{mathtools}
\usepackage{subfigure}
\usepackage{alphalph}
\usepackage{morefloats}
\usepackage{comment}
\usepackage{ulem}

\colorlet{shadecolor}{gray!15}
\definecolor{greenLinks}{rgb}{0, 0.6, 0} 
\definecolor{blueLinks}{rgb}{0, 0, 0.6}
\definecolor{redLinks}{rgb}{0.6, 0, 0}
\definecolor{tempText}{rgb}{0.55, 0.10,0.67}
\definecolor{eprintLinks}{rgb}{0.4, 0.4, 0.4}
\definecolor{journalLinks}{rgb}{0.6, 0, 0}

\hypersetup{colorlinks=true,linkcolor=blueLinks,citecolor=blueLinks,urlcolor=blueLinks}

\def\21{$\mathrm{SU(2)_L \otimes U(1)_Y}$ }

\newcommand{\AddrIPN}{%
Departamento de F\'{\i}sica, Centro de Investigaci\'on
  y de Estudios Avanzados del IPN,\\ Apartado Postal 14-740 07000,
  Ciudad de Mexico, Mexico}
  
\newcommand{\AddrIPhT}{Institut de Physique Th{\'e}orique, Universit{\'e} Paris Saclay, CNRS, CEA, F-91191 Gif-sur-Yvette, France}  

\newcommand{\AddrIFIC}{Instituto de F\'isica Corpuscular (CSIC-Universitat de Val\`encia),\\Parc Cient\'ific UV, C/ Catedr\`atico Jos\'e Beltr\'an, 2, 46980 Paterna, Spain}

\begin{document}

\title{Constraining Non-Standard Interactions with Coherent Elastic Neutrino-Nucleus Scattering at the European Spallation Source}

\author{Sabya Sachi Chatterjee} \email{sabya-sachi.chatterjee@ipht.fr} \affiliation{\AddrIPhT}

\author{St{\'e}phane Lavignac} \email{stephane.lavignac@ipht.fr} \affiliation{\AddrIPhT}

\author{O. G. Miranda} \email{omar.miranda@cinvestav.mx} \affiliation{\AddrIPN}

\author{G. Sanchez Garcia} \email{gsanchez@fis.cinvestav.mx} \affiliation{\AddrIPN}\affiliation{\AddrIFIC}

\begin{abstract}
\vspace{1cm}
The European Spallation Source (ESS), currently under construction in Sweden, will provide an intense pulsed neutrino flux allowing for high-statistics measurements of coherent elastic neutrino-nucleus scattering (CE$\nu$NS)
with advanced nuclear recoil detectors.
In this paper, we investigate in detail the possibility of constraining non-standard neutrino interactions (NSIs) through such precision CE$\nu$NS measurements at the ESS, considering the different proposed detection technologies, either alone or in combination.
We first study the sensitivity to neutral-current NSI parameters that each detector can reach in 3 years of data taking.
We then show that operating two detectors simultaneously
can significantly improve the expected sensitivity on flavor-diagonal NSI parameters.
Combining the results of two detectors turns out to be even more useful when two NSI parameters are assumed to be nonvanishing at a time. In this case, suitably chosen detector combinations can reduce the degeneracies between some pairs of NSI parameters to a small region of the parameter space.
\end{abstract}

\maketitle

\noindent

\section{Introduction}
\label{sec:intro}
Coherent elastic neutrino-nucleus scattering (CE$\nu$NS) was predicted
by Freedman more than forty years ago~\cite{PhysRevD.9.1389}.  In
this process, the neutrino interacts with the whole nucleus through
the exchange of a virtual $Z$ boson.
For this to happen, the transferred momentum should be comparable with the
inverse of the nucleus radius or smaller.
The contributions of the
individual nucleons then add coherently, resulting in a cross-section
that approximately scales like the squared number of neutrons in the
target nucleus, thus making CE$\nu$NS dominant over other neutrino
interactions, such as neutrino-electron scattering. Since the process is elastic, its only effect is to
redistribute the kinetic energy between the neutrino and the nucleus.
Therefore, its detection requires the measurement of the recoil energy
acquired by the nucleus. The smallness of the typical recoil energies
made the CE$\nu$NS observation a challenging task for decades. It
was only in 2017 that the COHERENT
collaboration~\cite{COHERENT:2015mry} reported the first measurement
of CE$\nu$NS~\cite{Akimov:2017ade}, using a CsI\,[Na]-based detector
and neutrinos produced from the Spallation Neutron Source (SNS) at Oak
Ridge National Laboratory. In the following years, the 
collaboration reported a second measurement in a LAr
detector~\cite{COHERENT:2020iec}, and another data set from the CsI
measurement was released~\cite{Akimov:2021dab}.

The COHERENT data has been used by several authors to perform tests of
the Standard Model (SM) and to constrain its possible extensions. As
a probe of standard physics, CE$\nu$NS is sensitive to the value of the
weak mixing angle at low energy~\cite{Papoulias:2019txv,Khan:2019cvi,Cadeddu:2019eta,Miranda:2020tif,Cadeddu:2021ijh}
and to nuclear parameters such as the neutron \textit{rms}
radius~\cite{Cadeddu:2017etk,Cadeddu:2019eta,Canas:2019fjw,Coloma:2020nhf}, which is well known only
for a few elements. As for physics beyond the SM, the COHERENT data
has been used to constrain non-standard interactions (NSIs) of
neutrinos~\cite{Barranco:2005yy,Scholberg:2005qs,Liao:2017uzy,Giunti:2019xpr,Miranda:2020tif,Denton:2020hop,Galindo-Uribarri:2020huw,Khan:2021wzy},
generalized neutrino interactions~\cite{Dutta:2015nlo,Lindner:2016wff,AristizabalSierra:2018eqm,Flores:2021kzl}, light mediators~\cite{Farzan:2018gtr,Denton:2018xmq,Flores:2020lji,Cadeddu:2020nbr,Banerjee:2021laz,delaVega:2021wpx,Bertuzzo:2021opb}, massive scalar and vector mediators~\cite{Billard:2018jnl,Arcadi:2019uif}, sterile neutrinos~\cite{Dutta:2015nlo,Kosmas:2017zbh,Miranda:2019skf,Miranda:2020syh,Khan:2022bcl},
non-unitarity of the lepton mixing matrix~\cite{Miranda:2020syh}, and neutrino electromagnetic properties such as the neutrino charge~\cite{Cadeddu:2019eta,Cadeddu:2020lky} and magnetic dipole moments~\cite{Miranda:2019wdy,Cadeddu:2020lky}. The precise measurement of the
CE$\nu$NS cross-section has applications in other fields, such as
direct dark matter detection, where it is important
for estimating the so-called neutrino
floor~\cite{AristizabalSierra:2021kht}. There is also an interest in
using CE$\nu$NS for nuclear security and monitoring
purposes~\cite{vonRaesfeld:2021gxl}.

Significant progress has been made in the characterization of detection materials for CE$\nu$NS, and several experimental programs aim at measuring CE$\nu$NS with neutrinos produced at spallation neutron sources in the coming years. For instance, the complete program of the COHERENT collaboration considers Ge and NaI detectors~\cite{COHERENT:2015mry}. Other experiments like the Coherent Captain Mills detector at Los Alamos National Laboratory will use LAr as target material~\cite{CCM:2021leg}. Another interesting proposal is to perform high-statistics CE$\nu$NS measurements at the upcoming European Spallation Source (ESS)~\cite{Baxter:2019mcx}.
Other experiments aim to measure CE$\nu$NS with reactor neutrinos, such as CONUS~\cite{CONUS:2021dwh}, CONNIE~\cite{CONNIE:2016nav},  NUCLEUS~\cite{NUCLEUS:2019igx} and RICOCHET~\cite{Billard:2016giu}.

The European Spallation Source (ESS), currently under construction in Lund, Sweden, will provide the world's most intense neutron beams, allowing for a rich experimental program. As a by-product, it will yield a large pulsed neutrino flux suitable for high-statistics measurements of coherent elastic neutrino-nucleus scattering (CE$\nu$NS), for which six different advanced detection technologies have been proposed in Ref.~\cite{Baxter:2019mcx}, namely CsI, Ge, Ar, Si, Xe and $\rm C_3F_8$ based detectors.
The possibility to probe NSIs through CE$\nu$NS measurements at the ESS was considered for the first time in the same paper~\cite{Baxter:2019mcx},
where the 90\% C.L. expected sensitivity to the flavor-diagonal non-standard neutrino couplings to the up quark was computed for the six above-mentioned detectors.
It was also shown that combining the data of a gas time projection chamber (TPC) detector filled alternatively with Xe and Ar would significantly reduce the degeneracy between the same two NSI parameters.

In this paper, we investigate in greater detail the capability of CE$\nu$NS measurements at the ESS to constrain non-standard neutrino interactions. We present the expected sensitivities of each of the six detectors proposed in Ref.~\cite{Baxter:2019mcx} to all relevant NSI parameters, considered one at a time or in pairs, and quantify the improvement that can be obtained by combining the data from two detectors. We also compare the capabilities of different detector combinations to break the degeneracies that appear when two NSI parameters are assumed to be simultaneously nonvanishing, and identify the most efficient ones for that purpose.

The paper is structured as follows: in Section~\ref{sec:theor-fram}, we give the cross-section for CE$\nu$NS both within the SM and in the presence of NSIs. In Section~\ref{sec:experimental-setup}, we present the experimental setup of the ESS, and we describe the procedure that we follow to study the expected sensitivities of the different proposed detectors to NSI parameters. We present our results in Section~\ref{sec:results}, and give our conclusions in Section~\ref{sec:conclusions}.

\section{Theoretical framework}
\label{sec:theor-fram}
At a spallation neutron source like the ESS, high-energy protons hit a fixed mercury target, producing neutrons and charged pions. The $\pi^-$ are absorbed by the nuclei of the target, whereas the $\pi^+$ lose their energy and decay at rest as $\pi^+ \rightarrow \mu^+ + \nu_{\mu}$. The muon neutrinos arising from this decay are monochromatic and are called \textit{prompt neutrinos}. The positively charged muons produced along with the muon neutrinos decay in turn at rest, producing two additional neutrinos, $\bar{\nu}_{\mu}$ and $\nu_e$, called \textit{delayed neutrinos}.
The total neutrino flux is therefore the sum of three components:
\begin{eqnarray}
\frac{dN_{\nu_\mu}}{dE_{\nu}} &= &\xi\, \delta\left(E_{\nu} - \frac{m_{\pi^+}^2-m_{\mu}^2}{2 m_{\pi^+}} \right), \\
\frac{dN_{\bar{\nu}_\mu}}{dE_{\nu}} &=& \xi \frac{64\, E_{\nu}^2}{m_{\mu}^3} \left(\frac{3}{4} - \frac{E_{\nu}}{m_{\mu}} \right), \\
\frac{dN_{\nu_e}}{dE_{\nu}} &=& \xi \frac{192\, E_{\nu}^2}{m_{\mu}^3}\left(\frac{1}{2} - \frac{E_{\nu}}{m_{\mu}} \right), 
\end{eqnarray}
where $E_{\nu}$ is the incoming neutrino energy (with $E_{\nu} \leq m_{\mu}/2 \simeq 52.8\, \rm MeV$ for the $\bar \nu_\mu$ and $\nu_e$ neutrinos) and $\xi = r N_{\rm POT}/4\pi L^2$, in which $r = 0.3$ is the number of neutrinos of each flavor produced by a single proton on target, $N_{\rm POT}$ is the total number of protons on target (POT) and $L$ is the distance from the source to the detector (with $N_{\rm POT} = 2.8 \times 10^{23}$ per calendar year and $L = 20\, \rm m$ at the ESS).

After traveling a distance $L$, neutrinos are detected through their
coherent elastic scattering off the target nuclei.
At tree level, in the SM, the differential cross-section for this process
is independent of the flavor of the incoming neutrino~\cite{Tomalak:2020zfh}.
For a neutrino (or antineutrino) of energy $E_{\nu}$ scattering off a nucleus with $Z$ protons,
$N$ neutrons and mass $M$, it is given by
\begin{eqnarray}
\label{Eq:crsection}
\frac{\mathrm{d}\sigma}{\mathrm{d}T} (E_{\nu}, T) = \frac{G_F^2M}{\pi}\left(1-\frac{MT}{2E_{\nu}^2}\right)\,F(|\vec{q}|^2)\,\left(Q_{W}^V\right)^2,
\end{eqnarray}
where $G_F$ is the Fermi constant, $T$ the recoil energy of the nucleus and $F(|\vec{q}|^2)$ the nuclear form factor (assumed to be the same for neutrons and protons) evaluated at the transferred three-momentum, $|\vec{q}| \simeq \sqrt{2MT}$.
The weak nuclear charge $Q_{W}^V$ reads, in the SM,
\begin{eqnarray}
\left(Q_{W,\rm{SM}}^V\right)^2 & = \left(Z\,g_V^p + N\,g_V^n \right)^2,
\label{eq:weak_charge_SM}
\end{eqnarray}
where $g_V^p = \frac{1}{2}-2\sin^2\theta_W$ and $g_V^n = -\frac{1}{2}$
 are the neutral current vector couplings\footnote{The contribution of
 axial couplings is proportional to the nucleus spin and is expected
 to be suppressed, with respect to the contributions of the vector
 couplings, by a factor $1/A$, where $A$ is the number of nucleons.}.  In our
 calculation, the weak mixing angle is taken at zero momentum
 transfer ($\sin^2\theta_W = 0.23867$~\cite{Erler:2017knj,PhysRevD.98.030001}),
 as the transferred three-momentum is constrained to be very small
 by the coherence condition (typically a few tens of MeV).
Throughout this work, we use the Helm
 parametrization~\cite{Helm:1956zz} for the nucleus form factor\footnote{For a detailed discussion on the impact of different form factors in CE$\nu$NS the reader can see, for example, Refs.~\cite{Papoulias:2018uzy,Papoulias:2019lfi,Hoferichter:2020osn}.}:
\begin{eqnarray}
F(|\vec{q}|^2) = 3\, \frac{j_1(|\vec{q}|R_0)}{|\vec{q}|R_0}\,e^{-|\vec{q}|^2s^2/2},
\end{eqnarray}
where $j_1$ denotes the spherical Bessel function of order one, $s = 0.9 \,  $fm~\cite{Friedrich:1982esq} is the surface thickness and $R_0^2 = 5/3 \left(R^2 - 3 s^2\right)$, with $R$ the neutron rms radius.

Let us now discuss how CE$\nu$NS is affected by the presence of
non-standard neutrino interactions (NSIs).  NSIs,
whose impact on neutrino propagation in matter was first discussed in Ref.~\cite{Wolfenstein:1977ue},  are
a possible low-energy manifestation of new physics
in the lepton sector. They
are usually parametrized in terms of 
higher-dimensional operators in the 
effective field theory of the SM below the weak scale.
As a neutral current process, CE$\nu$NS is only affected by neutral current NSIs (NC-NSIs),
which are described by
\begin{equation}
\mathcal{L}_{\mathrm{NC\mbox{-}NSI}} \;=\;
-2\sqrt{2}\,G_F\, 
\varepsilon_{\alpha\beta}^{fC}\,
\bigl(\overline{\nu}_\alpha\gamma^\mu P_L \nu_\beta\bigr)
\bigl(\overline{f}\gamma_\mu P_C f\bigr)
\;,
\label{H_NC-NSI}
\end{equation}
where $\alpha, \beta = e,\mu,\tau$ denote the neutrino flavors,  $f = e,u,d$ label the matter 
fermions, $C=L, R$ corresponds to the chirality of the matter fermion current, and $\varepsilon_{\alpha\beta}^{fC}$ 
are the strengths of the NSIs. Hermiticity of the Lagrangian requires $\varepsilon_{\beta\alpha}^{fC} = \left(\varepsilon_{\alpha\beta}^{fC}\right)^*$.
%
In this work, we will only consider the vector-type\footnote{NC-NSIs can also have axial components $\varepsilon_{\alpha\beta}^{fA} \equiv \varepsilon_{\alpha\beta}^{fR} - \varepsilon_{\alpha\beta}^{fL}$. However, their effect on CE$\nu$NS is negligible for heavy nuclei~\cite{Barranco:2005yy}.} NC-NSIs
\begin{eqnarray}
\varepsilon_{\alpha\beta}^{qV}\; \equiv\;  \varepsilon_{\alpha\beta}^{qL} + \varepsilon_{\alpha\beta}^{qR}\; ,
\qquad q = u, d\; .
\end{eqnarray}
Then, for an incident neutrino or antineutrino of flavor $\alpha$, the weak nuclear charge $Q_{W}^V$ in Eq.~(\ref{Eq:crsection}) is modified to~\cite{Barranco:2005yy}
\begin{eqnarray}
\left(Q_{W,\alpha}^V\right)^2 & = \left[Z\left(g_V^p + 2\varepsilon_{\alpha\alpha}^{uV}+\varepsilon_{\alpha\alpha}^{dV}\right) + N\left(g_V^n + \varepsilon_{\alpha\alpha}^{uV}+2\varepsilon_{\alpha\alpha}^{dV}\right)\right]^2 \nonumber \\
 & + \sum\limits_{\beta\neq \alpha} \left| Z\left(2 \varepsilon_{\alpha\beta}^{uV} + \varepsilon_{\alpha\beta}^{dV} \right)  + N\left(\varepsilon_{\alpha\beta}^{uV} + 2\varepsilon_{\alpha\beta}^{dV} \right) \right|^2.
 \label{eq:weak_charge_NSIs}
\end{eqnarray}
In the next sections, we study the sensitivity to NSIs that the proposed
CE$\nu$NS experiments at the ESS  can reach. 
We discuss how the
combination of different detectors can help to remove the degeneracies
arising when two NSI parameters are assumed to be nonvanishing at a
time.

\section{Experimental setup and simulation}
\label{sec:experimental-setup}

In this section, we briefly review the experimental specifications of
the different detectors proposed for CE$\nu$NS measurements at the
ESS. Then we give a detailed description of our
simulation procedure.

\subsection{Experimental setup options}
\label{sec:experimental-setup-1}
The European Spallation Source is a highly
ambitious and promising multi-disciplinary research facility under
completion at Lund, Sweden. It will use a very powerful
superconducting proton linac to generate the world's most intense
pulsed neutron beams.  As a byproduct, it will also generate
high-intensity, low-energy neutrino fluxes suitable for the study of
coherent elastic neutrino-nucleus scattering (CE$\nu$NS). The ESS will
use a proton beam energy of $2 \,\rm GeV$ and a beam power of $5\, \rm
MW$, resulting in $2.8\times 10^{23}$ protons on target (POT) per
calendar year ($208$ effective days), which will lead to an increase
of the neutrino flux by one order of magnitude with respect to
the current configuration of the Spallation Neutron Source
(SNS)~\cite{COHERENT:2015mry} located at Oak Ridge National
Laboratory, USA.  Furthermore, it is expected that a future upgrade of
the proton linac will increase the proton energy to $3.6\, \rm GeV$.
Since the construction of the ESS is underway, it has been proposed to
set up a number of different detectors sensitive to very low-energy
nuclear recoils, in order to take advantage of
the high-intensity neutrino flux to make precision CE$\nu$NS
measurements. Following the proposal of Ref.~\cite{Baxter:2019mcx}, we investigate
in this work the physics potential of six different detector targets
(CsI, Xe, Ge, Ar, Si, and $\rm C_3F_8$), alone or in combination.

The differential event rate $\mathrm{d}N / \mathrm{d}T$ for these detectors is
given by the convolution of the ESS flux and the CE$\nu$NS
cross-section $\mathrm{d} \sigma / \mathrm{d} T (E_{\nu},T)$
(with the weak nuclear charge given by Eq.~(\ref{eq:weak_charge_SM}) in the SM and by
Eq.~(\ref{eq:weak_charge_NSIs}) in the presence of NSIs) as a function
of the true nuclear recoil energy $T$:
\begin{eqnarray}
\frac{\mathrm{d}N}{\mathrm{d}T}\; =\; \mathcal{N}\!\!  \sum\limits_{\nu = \nu_e, \nu_{\mu}, \bar \nu_{\mu}}  \int_{E_{min}(T)}^{E_{max}}dE_{\nu}\;  \frac{\mathrm{d} N_{\nu}}{\mathrm{d} E_{\nu}}(E_{\nu})\: \frac{\mathrm{d} \sigma }{\mathrm{d} T}(E_{\nu},T)\; ,
\label{eq:events}
\end{eqnarray} 
where $\mathcal{N} = N_A M_{\rm det} / M_{\rm mol}$ is the number
of nuclei in the detector (with $N_A$ the Avogadro constant, $M_{\rm
det}$ the detector mass and $M_{\rm mol}$ the molar mass of the
detector material),
$E_{min}(T) = \sqrt{M\,T/2}$ (with $M$ the mass of the nucleus) is the
minimal neutrino energy required to induce a nuclear recoil energy
$T$, and $E_{max} = 52.8\, \rm MeV$ is the maximal energy that a
neutrino from the ESS flux can have. The maximal value of the nuclear
recoil energy is given by ${T}_{max} = 2E_{max}^2/M$.
We assume an experimental running time of $3$ years, which is quite
reasonable given the long-term physics program planned at the ESS. The
distance between the source and the detectors is taken to be
20\,$\rm{m}$. We make the conservative assumption of a detector signal
acceptance of $80\%$ above the nuclear recoil energy threshold.

\begin{table}[t!]
\newcommand{\mc}[3]{\multicolumn{#1}{#2}{#3}}
\newcommand{\mr}[3]{\multirow{#1}{#2}{#3}}
\centering
\begin{tabular}{|c|c|c|c|c|c|c|c|c|}
\hline
Detector & Target & Detector mass & Steady-state & $E^{ee}_{th}$ & QF & $T_{th}$ & $\frac{\Delta E}{E} (\%)$ & $\sigma_0$\\ 
technology & nucleus & (kg) & background & (keV$_{\rm ee}$) & $(\%)$ & (keV$_{\rm nr}$) & ${\rm at}\,\, T_{th}$ & (keV) \\
\hline
\hline
Cryogenic scintillator & CsI              & 22.5       & 10\,ckkd         & 0.1             & 10         & 1         &   30 & 0.3   \\ 
\hline
High-pressure gaseous TPC & Xe                & 20         & 10\,ckkd         & 0.18            & 20         & 0.9       &  40 & 0.36     \\ 
\hline
Charge-coupled device  & Si           & 1          & 1\,ckkd          & 0.007           & 20         & 0.16      &  60  & 0.096   \\ 
\hline
p-type point contact HPGe & Ge            & 7          & 15\,ckkd         & 0.12            & 20         & 0.6       &  15  & 0.09  \\ 
\hline
Scintillating bubble chamber & Ar         & 10         & 0.1\,c/kg-day    & $-$             & $-$        & 0.1       &  40  & 0.04   \\
\hline
Standard bubble chamber & $\rm C_3F_8$    & 10         & 15\,c/kg-day     & $-$             & $-$        & 0.2       &  40   & 0.08  \\
\hline
\end{tabular}
\caption{Main properties of the detectors proposed for CE$\nu$NS measurements at the ESS: technology, target nucleus, detector mass, steady-state background, electron recoil energy threshold ($E^{ee}_{th}$), quenching factor (QF), nuclear recoil energy threshold ($T_{th}$), energy resolution at $T_{th}$ in percent ($\Delta E / E$) and in keV ($\sigma_0$). Backgrounds (which do not include the $4\times 10^{-2}$ reduction by the ESS duty factor) are given in counts per kg and day for bubble chambers, for which only the total event rates are considered, while they are given in counts per keV, kg and day (ckkd) for the other detectors ($\rm keV_{ee}$ for the semiconductors Si and Ge, and $\rm keV_{nr}$ for CsI and Xe). All the information shown in this table is taken from Ref.~\cite{Baxter:2019mcx}. More details about the detector characteristics, including the quenching factors, can be found in the same reference and in references therein.}
\label{detectors_table1}
\end{table}
\begin{table}[h!]
\centering
\begin{tabular}{|c|c|c|c|c|c|}
\hline
\textbf{Target nucleus} & \textbf{   Z   } & \textbf{   N   } & \textbf{ Z/N } & \textbf{M (a.m.u)} & \textbf{R (fm)} \\ 
\hline
Cs              & 55       & 78         & 0.71         & 132.91                 & 4.83 \\ 
\hline
I               & 53       & 74         & 0.72         & 126.90                  & 4.83 \\ 
\hline
Xe              & 54       & 78         & 0.69         &  131.29                 & 4.79 \\ 
\hline
Si              & 14       & 14         & 1        & 27.98                  & 3.12 \\ 
\hline
Ge              & 32       & 40         & 0.8         &   72.00              & 4.06 \\ 
\hline
Ar              & 18       & 22         & 0.81         & 39.95                  & 3.43 \\
\hline
C               & 6          & 6        & 1           & 12.01                  & 2.47 \\
\hline
F               & 9       & 10         & 0.9         & 19.00                  & 2.90 \\
\hline
\end{tabular}
\caption{List of the target nuclei used in the detectors of Table~\ref{detectors_table1} with their main properties: number of protons (Z) and neutrons (N), number of protons to neutrons ratio (Z/N), mass in atomic units (M) and neutron rms radius in fm (R).}
\label{detectors_table2}
\end{table}

The main properties of the different detectors
considered in our analysis have been compiled in Ref.~\cite{Baxter:2019mcx}; for convenience,
we reproduce this information in Table~\ref{detectors_table1}.
In Table~\ref{detectors_table2}, we quote the
physical properties and the neutron rms radii of the target
nuclei~\cite{Angeli:2013epw}.  For recoil energy-sensitive detectors (CsI, Xe,
Ge and Si), we use a Gaussian
energy smearing for event reconstruction with a width 
$\sigma\left(T\right)\,=\, \sigma_0\sqrt{T/T_{th}}$, where $\sigma_0 =
(\Delta E/E) T_{th}$ is the energy resolution at the threshold (see
Table~\ref{detectors_table1}). Furthermore, following Ref.~\cite{Baxter:2019mcx}, we divide the range of
reconstructed nuclear recoil energies into bins, taking the width of
each bin to be twice the energy resolution at its
center\footnote{For instance, for the CsI detector,
the energy resolution at the center of the bin close to the peak of the event spectrum
shown in Fig.~1 is $\sigma (T_{peak}) = \sigma_0 \sqrt{T_{peak}/T_{th}} \simeq 0.62\, \rm keV$
(with $\sigma_0 = 0.3\, \rm keV$ and $T_{th} = 1\, {\rm keV}_{\rm nr}$ from Table~\ref{detectors_table1},
and $T_{peak} = 4.25\, {\rm keV}_{\rm nr}$),
corresponding to a bin width of $1.24\, {\rm keV}_{\rm nr}$.}, and compute the number of events bin by
bin, applying the energy smearing between the true and reconstructed
recoil energies and the detection signal acceptance mentioned
above. For bubble chamber detectors (Ar and C$_3$F$_8$), which lack
energy resolution, we compute the total number of events above nuclear
recoil energy threshold (assuming a
signal acceptance of $80\%$ as for the recoil energy-sensitive
detectors); no binning or energy smearing is needed in that case.

\subsection{Simulation procedure}
\label{sec:simulation-procedure}
In order to assess the statistical sensitivity to NSI parameters of the different detector setups at the ESS facility, we use the following $\chi^2$ function:
\begin{equation}
\chi^2(\kappa)\; =\; \underset{\xi}{min}\left[\sum\limits_{i}2\left\{N_i(\kappa, \xi) - \tilde{N_{i}} + \tilde{N_{i}}\,{\rm ln}\left(\frac{\tilde{N_{i}}}{N_i(\kappa, \xi)}\right) \right\} + \left(\frac{\xi_{sig}}{\sigma_{sig}}\right)^2 + \left(\frac{\xi_{bg}}{\sigma_{bg}}\right)^2 \right],
\label{eq:analysis}
\end{equation}
where $i$ labels the different energy bins, $\tilde{N_i}$ is the number of events in the $i$th bin predicted in the SM, and $N_i(\kappa, \xi)$ is the corresponding quantity in the presence of NSI parameters (collectively denoted by $\kappa$), including normalization factors, $\xi_{sig}$ and $\xi_{bg}$
, for the signal and background. Explicitly,
\begin{equation}
\tilde N_i = N^{sig}_i (SM) + N^{bg}_i\; ,  \qquad
N_i(\kappa, \xi) = N^{sig}_i (\kappa) (1+ \xi_{sig}) + N^{bg}_i (1 + \xi_{bg})\; ,
\label{eq:Ni}
\end{equation}
where $N^{sig}_i (SM)$ and $N^{sig}_i (\kappa)$ are the numbers of signal events in the $i$th bin predicted by the SM and by the SM extension including NSIs, respectively, and $N^{bg}_i$ is the number of background events in the same bin. The normalization factors, $\xi_{sig}$ and $\xi_{bg}$, quantify the systematic errors, with associated uncertainties $\sigma_{sig}$ and $\sigma_{bg}$. They are assumed to be the same for each bin, but independent of each other. Following Ref.~\cite{Baxter:2019mcx}, we assume $\sigma_{sig}$ and $\sigma_{bg}$ to be $10\%$ and $1\%$, respectively.

The main source of background events is assumed to be the steady-state background, which is dominated by cosmic ray interactions.  Beam-related backgrounds induced by the spallation neutrons and backgrounds from neutrino-induced neutrons have been neglected due to their small contributions. The expected level of steady-state backgrounds reported in Ref.~\citep{Baxter:2019mcx} is indicated in Table~\ref{detectors_table1} for each detection technology. For the $\rm CsI$ and $\rm Xe$ detectors, backgrounds are given in number of counts per keV$_{\rm nr}$, per kg per day, while for the two semiconductors, $\rm Si$ and $\rm Ge$, they are reported in number of counts per keV$_{\rm ee}$ per kg per day.  In the latter case, we use the quenching factors given in Table~\ref{detectors_table1} to convert the background value from keV$_{\rm ee}$ to keV$_{\rm nr}$.  For bubble chamber detectors, the backgrounds have been integrated above the nucleation threshold and are given in number of counts per kg and day.

\section{Numerical results}
\label{sec:results}

In this section, we present our results for the expected sensitivities of the detectors listed in Table~\ref{detectors_table1} to the NSI parameters $\varepsilon^{uV}_{\alpha \beta}$ and $\varepsilon^{dV}_{\alpha \beta}$ ($\alpha, \beta = e, \mu, \tau$) after 3 years of data taking at the ESS. Using Eq.~(\ref{eq:events}), we first compute the CE$\nu$NS signal rates predicted by the SM, as well as the expected background. Then we give the expected sensitivities of the different detectors to the NSI parameters taken one at a time, and quantify the improvement expected if two detectors operate simultaneously. Next we switch on a second NSI parameter, and show the future expected sensitivity of each detector in the corresponding two-dimensional parameter space. Finally, we show how using two detectors helps to partially break the degeneracies between different NSI parameters, and identify some of the most efficient detector combinations.

\subsection{Event Spectra}
\begin{figure}[t!]
\centering
\includegraphics[height=7.5cm, width=7.5cm]{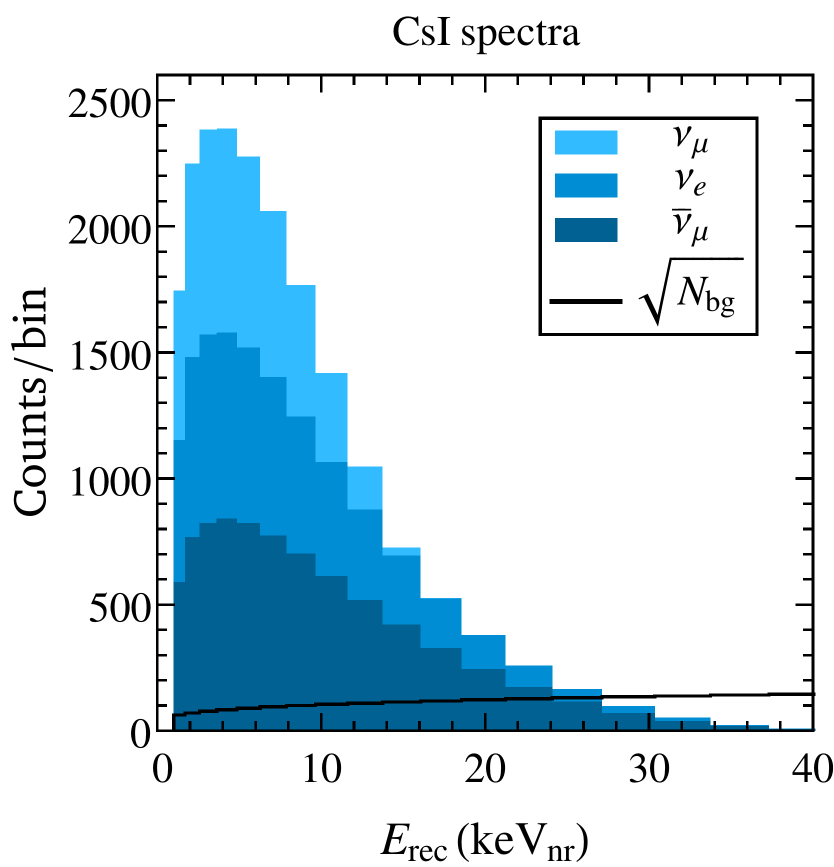}
\includegraphics[height=7.5cm, width=7.5cm]{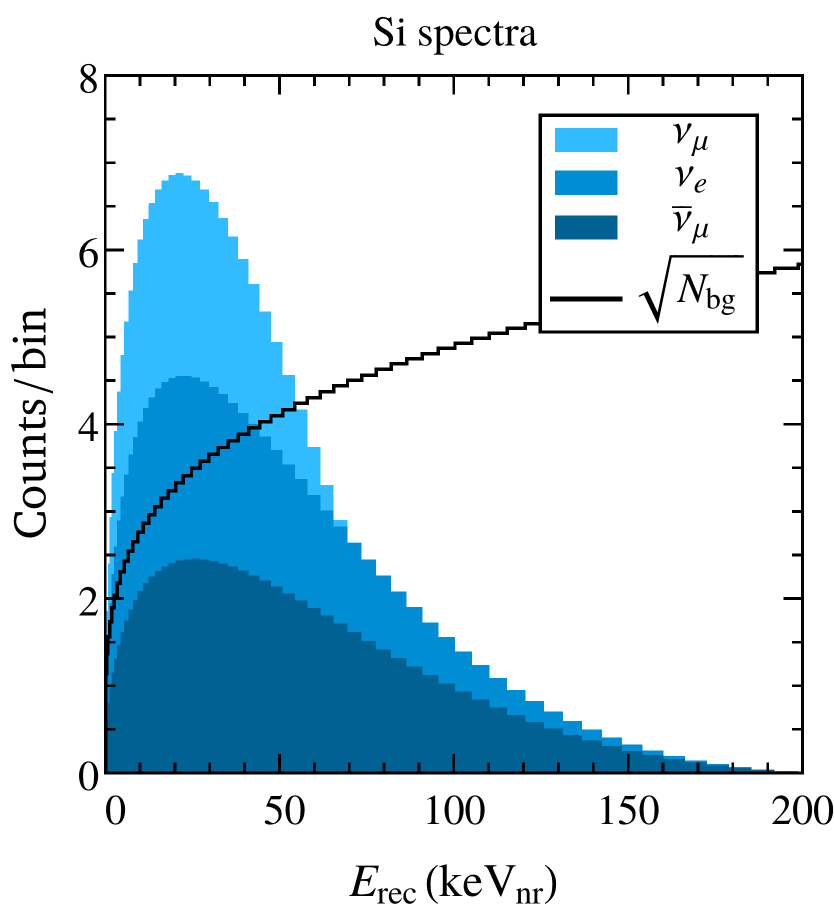}\\
\includegraphics[height=7.5cm, width=7.5cm]{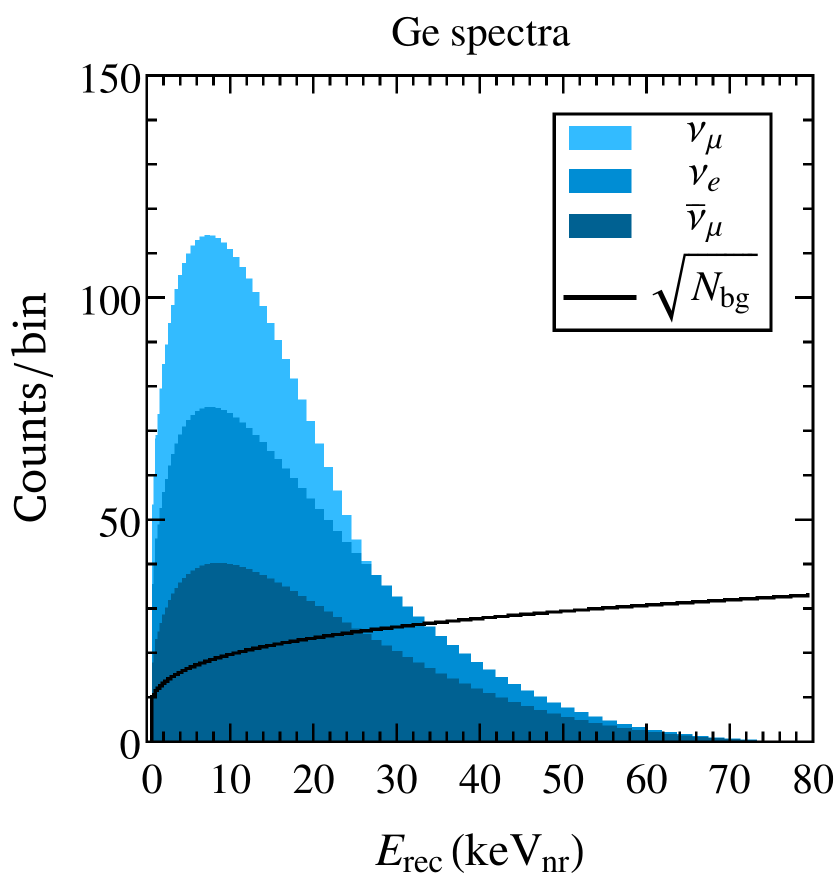}
\includegraphics[height=7.5cm, width=7.5cm]{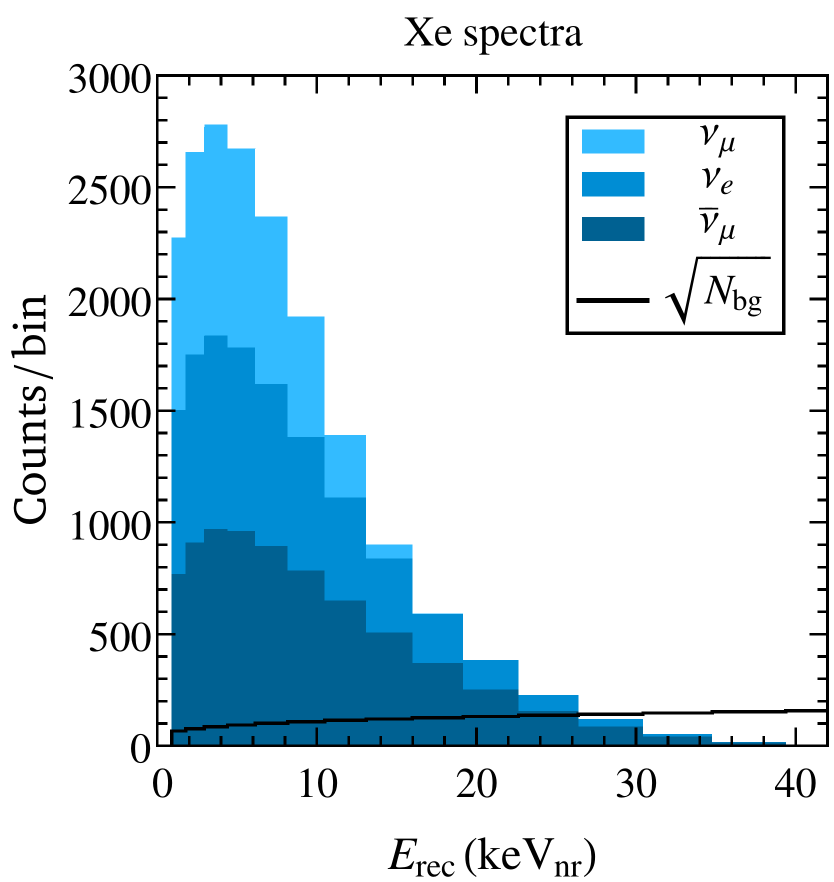}
\caption{Expected signal and background event spectra as a function of the reconstructed nuclear recoil energy for the $\rm CsI$, $\rm Si$, $\rm Ge$ and $\rm Xe$ detectors described in Table~\ref{detectors_table1}, assuming 3 years of data taking.  The signal events binning width is twice the energy resolution at its center, as explained in Section~\ref{sec:experimental-setup-1}, and the contributions from the different components of the ESS neutrino flux are distinguished by the various shades of the colored regions. The expected steady-state background is represented by the black curve, which for presentation purposes shows the square root of the actual number of events.}
\label{fig:spectra}
\end{figure}

In Fig.~\ref{fig:spectra}, we show the SM prediction for the CE$\nu$NS signal rate as a function of the reconstructed nuclear recoil energy for the $\rm CsI$, $\rm Si$, $\rm Ge$ and $\rm Xe$ detectors described in Table~\ref{detectors_table1}, assuming 3 years of data taking. The light, medium and dark blue histograms correspond to the contributions to the signal from the $\nu_{\mu}$, $\nu_e$ and $\bar{\nu}_{\mu}$ components of the ESS neutrino flux, respectively. The black curve represents the expected steady-state background as given in Table~\ref{detectors_table1}; for presentation purposes, the square root of the actual number of background events is shown.
Notice that the CsI event spectrum shown in Fig.~\ref{fig:spectra} perfectly agrees with the one displayed in Fig. 13 of Ref.~\cite{Baxter:2019mcx}.

Due to the long proton pulses of the ESS, it is impossible to distinguish experimentally between the contributions of the prompt and delayed components of the neutrino flux by using the additional timing information. However, above the maximal nuclear recoil energy that a prompt $\nu_{\mu}$ can induce, only the delayed $\nu_e$ and $\bar{\nu}_{\mu}$ contribute to the signal events. As an example, for the $\rm Si$ detector, the $\nu_{\mu}$ contribution sharply falls at the cut-off recoil energy $T^{max}_{prompt} = 2 E_{prompt}^2/M_{\rm Si} \simeq 68 \, {\rm keV}$, where $E_{prompt} = (m^2_{\pi^+} - m^2_\mu)/(2 m_{\pi^+}) \simeq 29.8\, {\rm MeV}$ is the energy of the monochromatic prompt neutrinos. A very small amount of $\nu_{\mu}$ events can still be observed just above $T^{max}_{prompt}$ due to the smearing by the energy resolution.

\begin{figure}[t!]
\centering
\includegraphics[width=1\textwidth]{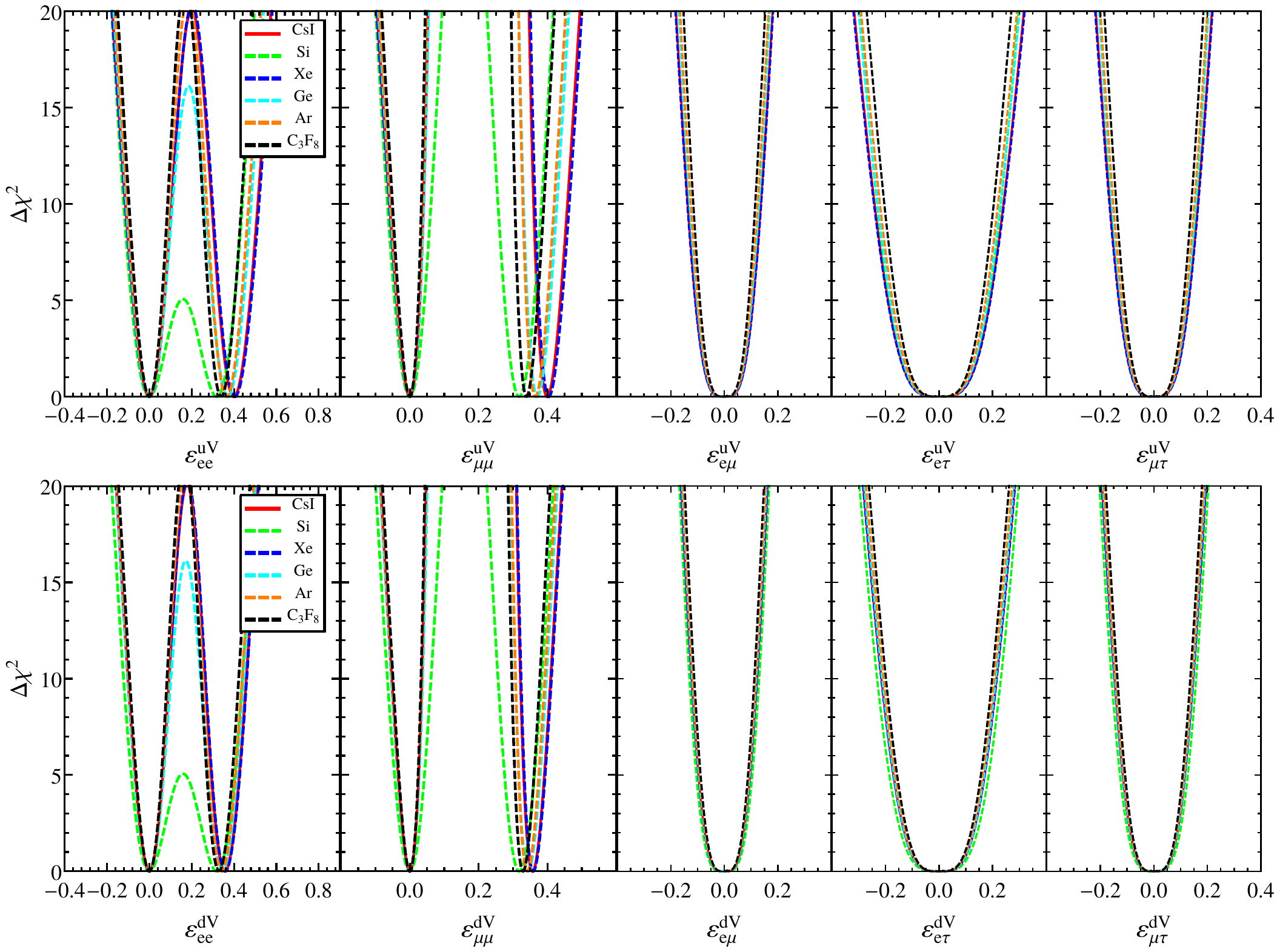}
\caption{One-dimensional projections of the expected sensitivities to the
NSI parameters $\varepsilon^{qV}_{ee}$, $\varepsilon^{qV}_{\mu\mu}$, $\varepsilon^{qV}_{e\mu}$, $\varepsilon^{qV}_{e\tau}$ and $\varepsilon^{qV}_{\mu\tau}$ (with $q=u$ for the upper panel and $q=d$ for the lower panel) of the detectors described in Table~\ref{detectors_table1}, assuming 3 years of data taking at the ESS. The red, green, blue, cyan, orange and black curves correspond to the $\rm CsI$, $\rm Si$, $\rm Xe$, $\rm Ge$, $\rm Ar$ and $\rm C_3F_8$ targets, respectively. All off-diagonal NSI parameters are assumed to be real.}
\label{fig:1D_proj}
\end{figure}
\begin{table}[t!]
\newcommand{\mc}[3]{\multicolumn{#1}{#2}{#3}}
\newcommand{\mr}[3]{\multirow{#1}{#2}{#3}}
\centering
\begin{tabular}{|c|c|c|c|c|c|}
\hline
 Target & Quark & \mc{2}{c|}{$\varepsilon^{qV}_{ee}$} &  \mc{2}{c|}{$\varepsilon^{qV}_{\mu\mu}$}  \\ 
 \cline{3-4}
 \cline{5-6}
 nucleus & type & $90\%\,\rm C.L.$  & $2\sigma\,\rm C.L.$ & $90\%\,\rm C.L.$ & $2\sigma\,\rm C.L.$    \\
\hline
\hline
 \mr{4}{*}{CsI}   &  \mr{2}{*}{$u$}  & $\left[-0.055, 0.051\right]\,  $  & $\left[-0.067, 0.064\right]\,  $       &  $\left[-0.027, 0.023\right]\, $     & $\left[-0.034, 0.028\right]\, $       \\ 
 &		& $\cup\;\left[0.35, 0.45\right]$   & $\cup\left[0.34, 0.47\right]$	& $\cup\left[0.38, 0.43\right]$	& $\cup\left[0.37, 0.43\right]$   \\
  & \mr{2}{*}{$d$}  & $\left[-0.040, 0.047\right]\, $ &$\left[-0.061, 0.057\right]\, $ &	  $\left[-0.024, 0.020\right]\, 	$  &  $\left[-0.031, 0.024\right]\,	$    \\
 &		& $\cup\left[0.31, 0.41\right]$	& $\cup\left[0.30, 0.42\right]$	&	$\cup\left[0.34, 0.38\right]$	&		$\cup\left[0.33, 0.39\right]	$	      \\ 
\hline
 \mr{4}{*}{Si}    &   \mr{2}{*}{$u$} & $\left[-0.063, 0.080\right]\, $  & $\left[-0.077, 0.109\right]\,  $ &  $\left[-0.032, 0.031\right]\, $  & $\left[-0.039, 0.037\right]\,  $     \\ 
 &		& $\cup\left[0.24, 0.38\right]$ & $\cup\left[0.21, 0.39\right]$	& $\cup\left[0.29, 0.35\right]$	&	$\cup\left[0.28, 0.36\right]	$   \\
  &  \mr{2}{*}{$d$} & $\left[-0.064, 0.08\right]\, $  & $\left[-0.077, 0.11\right]\, $ & 	$\left[-0.032, 0.031\right]\, 	$   & $\left[-0.039, 0.038\right]\, 	$     \\
  &		& $\cup\left[0.24, 0.38\right]$ &	$\cup\left[0.21, 0.39\right]$  	& $\cup\left[0.29, 0.35\right]$	&  $\cup\left[0.28, 0.36\right]	$    \\
\hline
 \mr{4}{*}{Xe}    &  \mr{2}{*}{$u$}  & $\left[-0.055, 0.052\right]\,  $  & $\left[-0.068, 0.063\right]\, $  & $\left[-0.027, 0.023\right]\,$  & $\left[-0.033, 0.026\right]	\, $              \\ 
 &		&  $\cup\left[0.35, 0.46\right]$  & $\cup\left[0.34, 0.47\right]$	& $\cup\left[0.38, 0.43\right]$	&    $\cup\left[0.38, 0.44\right]	$	 \\
  & \mr{2}{*}{$d$}  & $\left[-0.049, 0.047\right]\, $	& $\left[-0.061, 0.057\right]\, $  &  $\left[-0.025, 0.020\right]\, 	$ & $\left[-0.03, 0.023\right]\, 	$       \\
  &		& $\cup\left[0.31, 0.41\right]$	&	$\cup\left[0.30, 0.42\right]$  	& $\cup\left[0.34, 0.38\right]$	&   $\cup\left[0.34, 0.39\right]	$	 \\
\hline
 \mr{4}{*}{Ge}    &  \mr{2}{*}{$u$}   & $\left[-0.054, 0.054\right]\,  $  & $\left[-0.066, 0.064\right]\, $ &  $\left[-0.026, 0.022\right]\, $  & $\left[-0.032, 0.026\right]\, 	$        \\ 
 &		& $\cup\left[0.32, 0.42\right]$		&  $\cup\left[0.30, 0.44\right]$	& $\cup\left[0.35, 0.40\right]$ 	&	$\cup\left[0.34, 0.40\right]	$    	 \\
  & \mr{2}{*}{$d$}  & $\left[-0.05, 0.049\right]\, $		&  $\left[-0.061, 0.06\right]\, $ & $\left[-0.025, 0.021\right]\, 	$   & $\left[-0.029, 0.025\right]\, $     \\
  &		& $\cup\left[0.30, 0.39\right]$	  &  $\cup\left[0.28, 0.40\right]$  & $\cup\left[0.32, 0.37\right]$   	&	$\cup\left[0.32, 0.37\right]$	 		 \\
\hline
 \mr{4}{*}{Ar}    &  \mr{2}{*}{$u$}   & $\left[-0.05, 0.047\right]\,  $   & $\left[-0.062, 0.058\right]\, $ & $\left[-0.025, 0.021\right]\, $ & $\left[-0.031, 0.024\right]\, $           \\
 &		& 	$\cup\left[0.32, 0.42\right]$	&   $\cup\left[0.31, 0.43\right]$	 & $\cup\left[0.34, 0.39\right]$ 	&	$\cup\left[0.34, 0.40\right]	$			 \\
  & \mr{2}{*}{$d$}  & $\left[-0.047, 0.044\right]\, $		&  $\left[-0.058, 0.054\right]\, $ & $\left[-0.024, 0.019\right]\, 	$     & $\left[-0.029, 0.023\right]\, $     \\
  &		& $\cup\left[0.30, 0.39\right]$	 &     $\cup\left[0.29, 0.40\right]$	 &  $\cup\left[0.32, 0.36\right]$ 	&	$\cup\left[0.32, 0.37\right]$	      \\
\hline
 \mr{4}{*}{$\rm C_3F_8$} &  \mr{2}{*}{$u$} &   $\left[-0.046, 0.043\right]\,  $  & $\left[-0.057, 0.053\right]\, $  & $\left[-0.023, 0.019\right]\, $    & $\left[-0.029, 0.023\right]	\, $       \\
 &		& $\cup\left[0.30, 0.39\right]$		&  $\cup\left[0.29, 0.40\right]$	& $\cup\left[0.32, 0.36\right]$ 	&	$\cup\left[0.32, 0.37\right]$ 	 \\
  & \mr{2}{*}{$d$}  & $\left[-0.045, 0.042\right]\, $	&	$\left[-0.055, 0.052\right]\,  $ & $\left[-0.023, 0.018\right]\, 	$  & $\left[-0.028, 0.023\right]\, $    \\
  &		& $\cup\left[0.29, 0.37\right]$	&	$\cup\left[0.28, 0.39\right]$ &	$\cup\left[0.31, 0.35\right]$	&	$\cup\left[0.31, 0.36\right]$	 \\
\hline
\end{tabular}
\caption{Expected sensitivities of the detectors described in Table~\ref{detectors_table1} to the diagonal NSI parameters $\varepsilon^{qV}_{ee}$ and $\varepsilon^{qV}_{\mu\mu}$ ($q=u,d$), at the $90\%$ and $2\sigma$ confidence levels with one d.o.f. (i.e. $\Delta \chi^2 \leq 2.71$ and $\Delta \chi^2 \leq 4$, respectively), corresponding to the one-dimensional projections shown in Fig.~\ref{fig:1D_proj}. Experimental running time: 3 years.}
\label{const_table2}
\end{table}
\begin{table}[t!]
\newcommand{\mc}[3]{\multicolumn{#1}{#2}{#3}}
\newcommand{\mr}[3]{\multirow{#1}{#2}{#3}}
\centering
\begin{tabular}{|c|c|c|c|c|c|c|c|}
\hline
 Target & Quark & \mc{2}{c|}{$\varepsilon^{qV}_{e\mu}$} &  \mc{2}{c|}{$\varepsilon^{qV}_{e\tau}$} &\mc{2}{c|}{$\varepsilon^{qV}_{\mu\tau}$}  \\ 
 \cline{3-4}
 \cline{5-6}
 \cline{7-8}
 nucleus & type & $90\%\,\rm C.L.$  & $2\sigma\,\rm C.L.$ & $90\%\,\rm C.L.$ & $2\sigma\,\rm C.L.$  & $90\%\,\rm C.L.$ & $2\sigma\,\rm C.L.$   \\
\hline
\hline
 \mr{4}{*}{CsI}   &  \mr{2}{*}{$u$}  &  \mr{2}{*}{$\left[-0.089, 0.089\right]$}   & \mr{2}{*}{$\left[-0.1, 0.1\right]$} & \mr{2}{*}{$\left[-0.16, 0.16\right]$} & \mr{2}{*}{$\left[-0.18, 0.18\right]$}  & \mr{2}{*}{$\left[-0.11, 0.11\right]$}   &     \mr{2}{*}{$\left[-0.12, 0.12\right]$} \\ 
 &		&     &  	&	  &     &     &  \\
  & \mr{2}{*}{$d$}  & \mr{2}{*}{$\left[-0.08, 0.08\right]$}   &\mr{2}{*}{$\left[-0.09, 0.09\right]$} & \mr{2}{*}{$\left[-0.14, 0.14\right]$} 	&  \mr{2}{*}{$\left[-0.16, 0.16\right]$}  & \mr{2}{*}{$\left[-0.096, 0.096\right]$}  & \mr{2}{*}{$\left[-0.11, 0.11\right]$}  \\
 &		&	& 	&		&			 &   &      \\ 
\hline
 \mr{4}{*}{Si}    &   \mr{2}{*}{$u$} & \mr{2}{*}{$\left[-0.088, 0.088\right]$}  & \mr{2}{*}{$\left[-0.099, 0.099\right]$}  & \mr{2}{*}{$\left[-0.15, 0.15\right]$} & \mr{2}{*}{$\left[-0.17, 0.17\right]$}  & \mr{2}{*}{$\left[-0.11, 0.11\right]$}    &\mr{2}{*}{$\left[-0.12, 0.12\right]$}    \\ 
 &		&     & 	&	   &	  &      &  \\
  &  \mr{2}{*}{$d$} & \mr{2}{*}{$\left[-0.088, 0.088\right]$}  & \mr{2}{*}{$\left[-0.098, 0.098\right]$} & \mr{2}{*}{$\left[-0.15, 0.15\right]$}  	& \mr{2}{*}{$\left[-0.17, 0.17\right]$} & \mr{2}{*}{$\left[-0.11, 0.11\right]$}   & \mr{2}{*}{$\left[-0.12, 0.12\right]$}  \\
  &		&	&		&	&   &   &  \\
\hline
 \mr{4}{*}{Xe}    &  \mr{2}{*}{$u$}  & \mr{2}{*}{$\left[-0.09, 0.09\right]$}  & \mr{2}{*}{$\left[-0.1, 0.1\right]$}    &  \mr{2}{*}{$\left[-0.16, 0.16\right]$}    & \mr{2}{*}{$\left[-0.18, 0.18\right]$}  & \mr{2}{*}{$\left[-0.11, 0.11\right]$}     & \mr{2}{*}{$\left[-0.12, 0.12\right]$}     \\ 
 &		&    & 		&	&	 &   &   \\
  & \mr{2}{*}{$d$}  & \mr{2}{*}{$\left[-0.08, 0.08\right]$} 	& \mr{2}{*}{$\left[-0.09, 0.09\right]$}   & \mr{2}{*}{$\left[-0.14, 0.14\right]$}    & \mr{2}{*}{$\left[-0.16, 0.16\right]$}  & \mr{2}{*}{$\left[-0.097, 0.097\right]$}    & \mr{2}{*}{$\left[-0.11, 0.11\right]$}     \\
  &		&	&		&	&	 &   &   \\
\hline
 \mr{4}{*}{Ge}    &  \mr{2}{*}{$u$}   & \mr{2}{*}{$\left[-0.085, 0.085\right]$} 	& \mr{2}{*}{$\left[-0.096, 0.096\right]$}   & \mr{2}{*}{$\left[-0.15, 0.15\right]$}  & \mr{2}{*}{$\left[-0.17, 0.17\right]$}  &  \mr{2}{*}{$\left[-0.1, 0.1\right]$}     & \mr{2}{*}{$\left[-0.11, 0.11\right]$}     \\ 
 &		& 		&  		&  		&	   &   		&  	 \\
  & \mr{2}{*}{$d$}  & \mr{2}{*}{$\left[-0.079, 0.079\right]$} &  \mr{2}{*}{$\left[-0.088, 0.088\right]$} & \mr{2}{*}{$\left[-0.14, 0.14\right]$}  & \mr{2}{*}{$\left[-0.16, 0.16\right]$}   & \mr{2}{*}{$\left[-0.095, 0.095\right]$}  & \mr{2}{*}{$\left[-0.11, 0.11\right]$}  \\
  &		&	  &    &    	&		  &   &	 \\
\hline
 \mr{4}{*}{Ar}    &  \mr{2}{*}{$u$}   & \mr{2}{*}{$\left[-0.082, 0.082\right]$}    &\mr{2}{*}{$\left[-0.092, 0.092\right]$} & \mr{2}{*}{$\left[-0.14, 0.14\right]$}   & \mr{2}{*}{$\left[-0.16, 0.16\right]$}   & \mr{2}{*}{$\left[-0.099, 0.099\right]$}   & \mr{2}{*}{$\left[-0.11, 0.11\right]$}   \\
 &		& 		&   	 &  	&	 &   &		 \\
  & \mr{2}{*}{$d$}  & \mr{2}{*}{$\left[-0.077, 0.077\right]$}		&  \mr{2}{*}{$\left[-0.086, 0.086\right]$} &  \mr{2}{*}{$\left[-0.13, 0.13\right]$}    & \mr{2}{*}{$\left[-0.15, 0.15\right]$}   & \mr{2}{*}{$\left[-0.093, 0.093\right]$}  & \mr{2}{*}{$\left[-0.1, 0.1\right]$}   \\
  &		&	  &   	 &   	&	  &   	&     \\
\hline
 \mr{4}{*}{$\rm C_3F_8$} &  \mr{2}{*}{$u$} & \mr{2}{*}{$\left[-0.075, 0.075\right]$}	& \mr{2}{*}{$\left[-0.085, 0.085\right]$} &  \mr{2}{*}{$\left[-0.13, 0.13\right]$} & \mr{2}{*}{$\left[-0.15, 0.15\right]$}    & \mr{2}{*}{$\left[-0.092, 0.092\right]$}  & \mr{2}{*}{$\left[-0.1, 0.1\right]$}     \\
 &		& 		&   	&	  &		  &   &	 \\
  & \mr{2}{*}{$d$}  & \mr{2}{*}{$\left[-0.073, 0.073\right]$}	&	\mr{2}{*}{$\left[-0.083, 0.083\right]$} & \mr{2}{*}{$\left[-0.13, 0.13\right]$}  & \mr{2}{*}{$\left[-0.15, 0.15\right]$}   & \mr{2}{*}{$\left[-0.089, 0.089\right]$}  & \mr{2}{*}{$\left[-0.1, 0.1\right]$}  \\
  &		&	&	 &		&	   &   &	\\
\hline
\end{tabular}
\caption{Expected sensitivities of the detectors described in Table~\ref{detectors_table1} to the off-diagonal NSI parameters $\varepsilon^{qV}_{e\mu}$, $\varepsilon^{qV}_{e\tau}$ and $\varepsilon^{qV}_{\mu\tau}$ ($q=u,d$), at the $90\%$ and $2\sigma$ confidence levels with one d.o.f. (i.e. $\Delta \chi^2 \leq 2.71$ and $\Delta \chi^2 \leq 4$, respectively), corresponding to the one-dimensional projections shown in Fig.~\ref{fig:1D_proj}. Experimental running time: 3 years. All NSI parameters are assumed to be real.}
\label{const_table3}
\end{table}
\subsection{Detector sensitivities to NSI parameters}
We are now ready to discuss the expected sensitivities to NSI parameters of the detectors described in Table~\ref{detectors_table1}.
By definition, CE$\nu$NS can only probe NC-NSIs involving quarks, parametrized by the coefficients $\varepsilon^{uV}_{\alpha \beta}$ and $\varepsilon^{dV}_{\alpha \beta}$ ($\alpha, \beta = e, \mu, \tau$). Furthermore, since the ESS neutrino beam contains electron and muon (anti)neutrinos, but no tau neutrinos, only four out of six flavor-diagonal parameters can be constrained (namely, $\varepsilon^{qV}_{ee}$ and $\varepsilon^{qV}_{\mu\mu}$, with $q=u$ or $d$). On the other hand, all six flavor off-diagonal parameters $\varepsilon^{qV}_{e\mu}$, $\varepsilon^{qV}_{e\tau}$ and $\varepsilon^{qV}_{\mu\tau}$ ($q=u,d$) contribute to the CE$\nu$NS signal events. Throughout this work, we assume the off-diagonal NSI parameters to be real (the flavor diagonal ones are always real due to the hermiticity of the Lagrangian). In our analysis, we assume the true hypothesis to be the Standard Model, and we test nonzero NSI parameters against it. We therefore take $\Delta\chi^2 = \chi^2(\kappa)$, with $\chi^2(\kappa)$ the function defined in Eq.~(\ref{eq:analysis}), and $\kappa$ the set of NSI parameters assumed to be nonvanishing.

We start by considering a single NSI parameter at a time. In Fig.~\ref{fig:1D_proj}, we show the associated $\Delta \chi^2$ functions computed for each of the six detectors described in Table~\ref{detectors_table1}. Each plot corresponds to a different nonvanishing NSI parameter (from left to right: $\varepsilon^{qV}_{ee}$, $\varepsilon^{qV}_{\mu\mu}$, $\varepsilon^{qV}_{e\mu}$, $\varepsilon^{qV}_{e\tau}$ and $\varepsilon^{qV}_{\mu\tau}$, with $q=u$ for the upper panel and $q=d$ for the lower panel). The red, green, blue, cyan, orange and black curves refer to the detectors with CsI, Si, Xe, Ge, Ar and $\rm C_3F_8$ targets, respectively. One can see that, individually, all these targets have similar sensitivities, except for Si, whose sensitivity to the diagonal NSI parameters is slightly degraded due to low statistics. However, due to its proton to neutron ratio, which significantly differs from the ones of the other target nuclei considered, the Si nucleus will have a significant impact when combining the results of different detectors, as we will see later. In Table~\ref{const_table2}, we give the expected sensitivity of each detector to the diagonal NSI parameters at the $90\%$ and $2\sigma$ confidence levels with one degree of freedom (d.o.f.), i.e. $\Delta \chi^2 \leq 2.71$ and $\Delta \chi^2 \leq 4$, respectively.
For each parameter, two disconnected regions of values appear, associated with the two minima of the $\Delta \chi^2$.  Finally, Table~\ref{const_table3} shows the expected sensitivities to the off-diagonal NSI parameters. In this case, there is a single region of expected allowed values for each parameter, centered around zero.
These expected sensitivities represent a substantial improvement with respect to the constraints extracted from the current COHERENT results~\cite{Giunti:2019xpr,Miranda:2020tif}.

\begin{figure}[t!]
\centering
\includegraphics[width=1\textwidth]{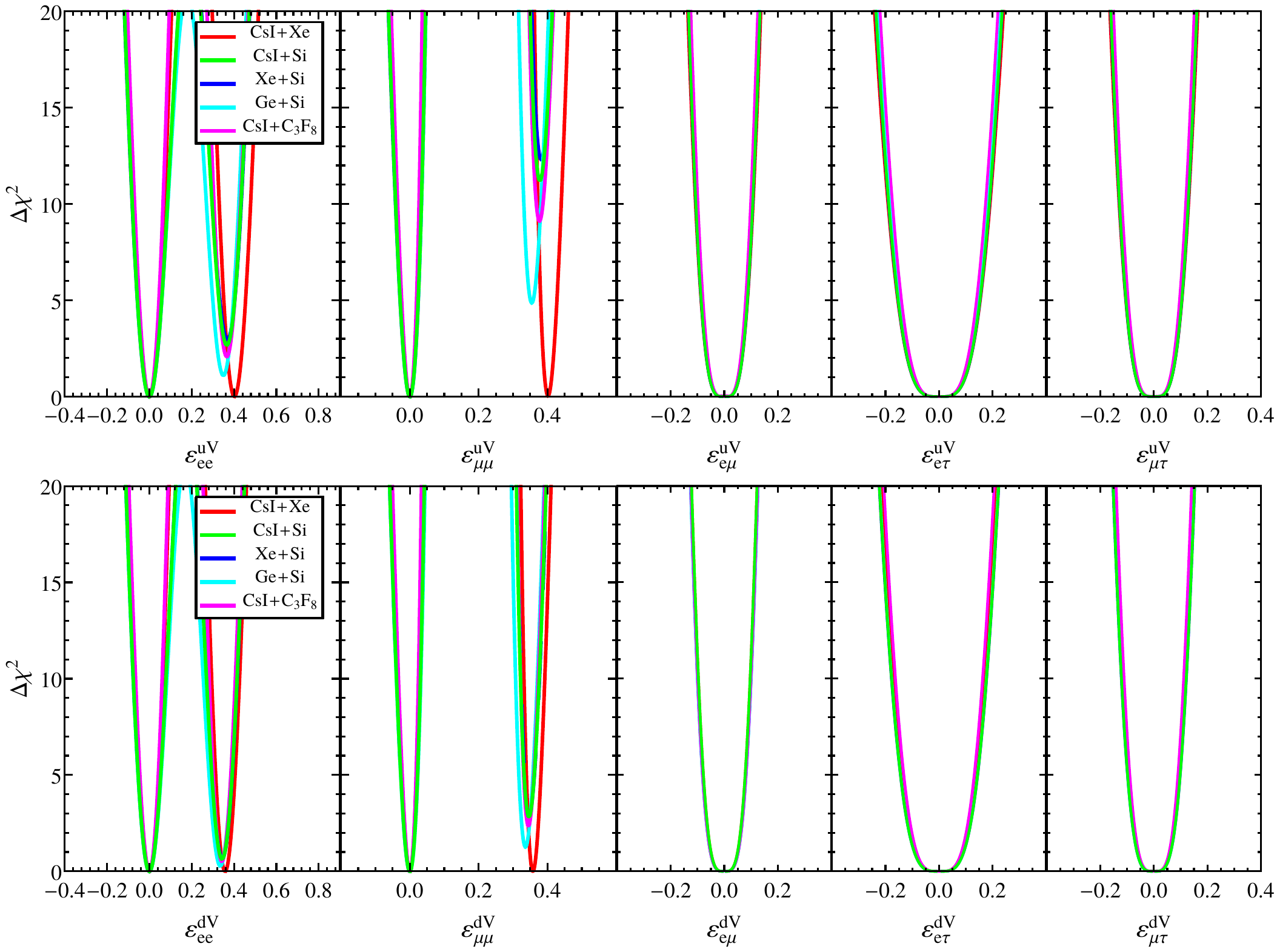}
\caption{One-dimensional projections of the expected sensitivities to the
NSI parameters $\varepsilon^{qV}_{ee}$, $\varepsilon^{qV}_{\mu\mu}$, $\varepsilon^{qV}_{e\mu}$, $\varepsilon^{qV}_{e\tau}$ and $\varepsilon^{qV}_{\mu\tau}$ (with $q=u$ for the upper panel and $q=d$ for the lower panel) of the detectors described in Table~\ref{detectors_table1}, assuming 3 years of data taking at the ESS. The red, green, blue, cyan and magenta curves correspond to the $\rm CsI + Xe$, $\rm CsI+Si$, $\rm Xe + Si$, $\rm Ge + Si$ and $\rm CsI+ C_3F_8$ target combinations, respectively. All off-diagonal NSI parameters are assumed to be real.}
\label{fig:1D_proj_combined}
\end{figure}

Let us now see how combining two detectors with different target materials can improve the sensitivity to NSI parameters. In Fig.~\ref{fig:1D_proj_combined}, we show the $\Delta \chi^2$ functions computed by combining the event spectra of two detectors. As in Fig.~\ref{fig:1D_proj}, each plot corresponds to a different nonvanishing NSI parameter (from left to right: $\varepsilon^{qV}_{ee}$, $\varepsilon^{qV}_{\mu\mu}$, $\varepsilon^{qV}_{e\mu}$, $\varepsilon^{qV}_{e\tau}$ and $\varepsilon^{qV}_{\mu\tau}$, with $q=u$ for the upper panel and $q=d$ for the lower panel). For simplicity, we consider only a few out of the many possible combinations between the six detectors, which we choose among the most efficient ones in terms of sensitivity improvement. The red, green, blue, cyan and magenta curves refer to the combinations $\rm CsI+Xe$, $\rm CsI+Si$, $\rm Xe+Si$, $\rm Ge+Si$ and $\rm CsI+C_3F_8$, respectively. As can be seen from these plots, the most significant sensitivity improvement with respect to the setup with a single detector is obtained from the combination $\rm CsI+Si$, especially for the parameters $\varepsilon^{uV}_{ee}$, $\varepsilon^{uV}_{\mu\mu}$ and $\varepsilon^{dV}_{\mu\mu}$. In Tables~\ref{const_table4} and~\ref{const_table5}, we give the expected sensitivities of this particular combination of detectors to the diagonal and off-diagonal NSI parameters, respectively. As can be seen by comparing these expected sensitivities with the ones appearing in Tables~\ref{const_table2} and~\ref{const_table3}, the improvement is significant for the off-diagonal parameters, and even more remarkable for some of the diagonal parameters. For instance, the expected allowed region around $0.3 - 0.4$ for $\varepsilon^{uV}_{\mu\mu}$ in Table~\ref{const_table2} can be excluded at more than $3\sigma$ by the combination $\rm CsI+Si$.
By contrast, the analogous regions for $\varepsilon^{uV}_{ee}$ and $\varepsilon^{dV}_{\mu\mu}$ can only be excluded at the $90\%$ confidence level.

%
%
\begin{table}[t!]
\newcommand{\mc}[3]{\multicolumn{#1}{#2}{#3}}
\newcommand{\mr}[3]{\multirow{#1}{#2}{#3}}
\centering
\begin{tabular}{|c|c|c|c|c|c|}
\hline
 Target & Quark & \mc{2}{c|}{$\varepsilon^{qV}_{ee}$} &  \mc{2}{c|}{$\varepsilon^{qV}_{\mu\mu}$}  \\ 
 \cline{3-4}
 \cline{5-6}
 nucleus & type & $90\%\,\rm C.L.$  & $2\sigma\,\rm C.L.$ & $90\%\,\rm C.L.$ & $2\sigma\,\rm C.L.$    \\
\hline
\hline
 \mr{4}{*}{CsI + Si}   &  \mr{2}{*}{$u$}  & $\left[-0.041, 0.042\right]\,$  & $\left[-0.05, 0.052\right]\,  $ &  $\left[-0.02, 0.016\right]\,$     & $\left[-0.025, 0.022\right]\,$       \\ 
 &		& $-$   & $\cup\left[0.34, 0.39\right]$	  &   $-$ 	& $-$   \\
  & \mr{2}{*}{$d$}  & $\left[-0.038, 0.039\right]\, $ &$\left[-0.046, 0.047\right]\, $ &	  $\left[-0.018, 0.015\right]\,	$  &  $\left[-0.023, 0.02\right]\, 	$    \\
 &		& $\cup\left[0.31, 0.38\right]$  	&   $\cup\left[0.30, 0.39\right]$	&	$-$	&		$\cup\left[0.34, 0.36\right]$	      \\ 
\hline
\end{tabular}
\caption{Same as Table~\ref{const_table2}, but for the combined analysis of the CsI and Si detectors,
shown in Fig.~\ref{fig:1D_proj_combined}.}
\label{const_table4}
\end{table}
\begin{table}[t!]
\newcommand{\mc}[3]{\multicolumn{#1}{#2}{#3}}
\newcommand{\mr}[3]{\multirow{#1}{#2}{#3}}
\centering
\begin{adjustbox}{width=1\textwidth}
\begin{tabular}{|c|c|c|c|c|c|c|c|}
\hline
 Target & Quark & \mc{2}{c|}{$\varepsilon^{qV}_{e\mu}$} &  \mc{2}{c|}{$\varepsilon^{qV}_{e\tau}$} &\mc{2}{c|}{$\varepsilon^{qV}_{\mu\tau}$}  \\ 
 \cline{3-4}
 \cline{5-6}
 \cline{7-8}
 nucleus & type & $90\%\,\rm C.L.$  & $2\sigma\,\rm C.L.$ & $90\%\,\rm C.L.$ & $2\sigma\,\rm C.L.$  & $90\%\,\rm C.L.$ & $2\sigma\,\rm C.L.$   \\
\hline
\hline
 \mr{4}{*}{CsI+Si}   &  \mr{2}{*}{$u$}  &  \mr{2}{*}{$\left[-0.074, 0.074\right]$}   & \mr{2}{*}{$\left[-0.081, 0.081\right]$} & \mr{2}{*}{$\left[-0.13, 0.13\right]$} & \mr{2}{*}{$\left[-0.14, 0.14\right]$}  & \mr{2}{*}{$\left[-0.09, 0.09\right]$}   &     \mr{2}{*}{$\left[-0.10, 0.10\right]$} \\ 
 &		&     &  	&	  &     &     &  \\
  & \mr{2}{*}{$d$}  & \mr{2}{*}{$\left[-0.068, 0.068\right]$}   &\mr{2}{*}{$\left[-0.075, 0.075\right]$} & \mr{2}{*}{$\left[-0.12, 0.12\right]$} 	&  \mr{2}{*}{$\left[-0.14, 0.14\right]$}  & \mr{2}{*}{$\left[-0.085, 0.085\right]$}  & \mr{2}{*}{$\left[-0.092, 0.092\right]$}  \\
 &		&	& 	&		&			 &   &      \\ 
\hline
\end{tabular}
\end{adjustbox}
\caption{Same as Table~\ref{const_table3}, but for the combined analysis of the CsI and Si detectors,
shown in Fig.~\ref{fig:1D_proj_combined}.}
\label{const_table5}
\end{table}

NC-NSI parameters are also constrained by neutrino oscillation
experiments, as they modify neutrino propagation in
matter~\cite{Wolfenstein:1977ue,Mikheyev:1985zog}. However, while
CE$\nu$NS is only sensitive to the parameters
$\epsilon^{uV}_{\alpha \beta}$ and $\epsilon^{dV}_{\alpha \beta}$,
oscillation data also depend on the neutrino couplings to
electrons, $\epsilon^{eV}_{\alpha \beta}$. More precisely, the Hamiltonian in matter depends on the
combinations of NSI parameters
$\epsilon^m_{\alpha \beta}(x) \equiv \epsilon^{eV}_{\alpha \beta}
+ \epsilon^{pV}_{\alpha \beta} + Y_n(x)\, \epsilon^{nV}_{\alpha \beta}$,
where $\epsilon^{pV}_{\alpha \beta} = 2 \epsilon^{uV}_{\alpha \beta}
+ \epsilon^{dV}_{\alpha \beta}$, $\epsilon^{nV}_{\alpha \beta}
= \epsilon^{uV}_{\alpha \beta} + 2 \epsilon^{dV}_{\alpha \beta}$, and the
$x$-dependence comes from the varying chemical composition of the
medium, $Y_n(x) = n_n(x) / n_p(x)$ (with $n_n(x)$ and $n_p(x)$ the
neutron and proton densities).
Furthermore, only the differences of diagonal NSI parameters
are constrained by oscillation data.
In addition to giving a more direct access to non-standard neutrino couplings to quarks,
CE$\nu$NS measurements are crucial
to discriminate between the LMA and the LMA-Dark explanations
of neutrino oscillation data~\cite{Miranda:2004nb,Coloma:2016gei}.
In Refs.~\cite{Esteban:2018ppq,Coloma:2019mbs,Chaves:2021pey},
a combined analysis of
oscillation data and of the COHERENT results was performed under the assumption
that all $\epsilon^e_{\alpha \beta}$ parameters vanish and that the
non-standard neutrino couplings to up and down quarks
are proportional to each other, $\epsilon^{uV}_{\alpha \beta} \propto \epsilon^{dV}_{\alpha \beta}$.
It should be kept in mind that the associated bounds on NSI parameters
do depend on these assumptions and cannot be directly compared with the expected sensitivities
from future CE$\nu$NS experiments at the ESS discussed in this paper.

\subsection{Breaking degeneracies in the presence of two NSI parameters}

We now move on to study the case where two NSI parameters are different from zero at a time. As can be expected, the sensitivity of a single detector is strongly limited by the existence of degeneracies between the two parameters.
These degeneracies are a consequence of the fact that (considering for simplicity a single-nucleus target) the number of CE$\nu$NS events in each energy bin depends only on two nucleus-dependent combinations of NSI parameters, $(Q_{W,e}^V)^2$ and $(Q_{W,\mu}^V)^2\,$, defined in Eq.~(\ref{eq:weak_charge_NSIs}). As a result, it is not possible to experimentally distinguish between different sets of NSI parameter values giving the same $(Q_{W,e}^V)^2$ and $(Q_{W,\mu}^V)^2\,$; this is called a degeneracy, and it results in extended allowed regions in two-parameter sensitivity plots obtained from a single nuclear target\footnote{In practice, degeneracies also arise between different sets of NSI parameter values giving the same total number of events, but different values for $(Q_{W,e}^V)^2$ and $(Q_{W,\mu}^V)^2\,$. This is a consequence of the experimental difficulty
to distinguish between the contribution of $\nu_e$ (which depends on $(Q_{W,e}^V)^2$) from the one of $\nu_\mu$ and $\bar \nu_\mu$ (which depends on $(Q_{W,\mu}^V)^2$). This is the case, for instance, of the degeneracies observed between the parameters $\varepsilon^{qV}_{ee}$ and $\varepsilon^{q'V}_{\mu\mu}$ ($q,q'=u,d$) in Fig.~\ref{fig:contour1}.}.
We will however see that combining the event spectra of two suitably chosen detectors can significantly improve the experimental sensitivity and partially break these degeneracies.

\begin{figure}[t]
\centering
\includegraphics[height=7.6cm, width=7.6cm]{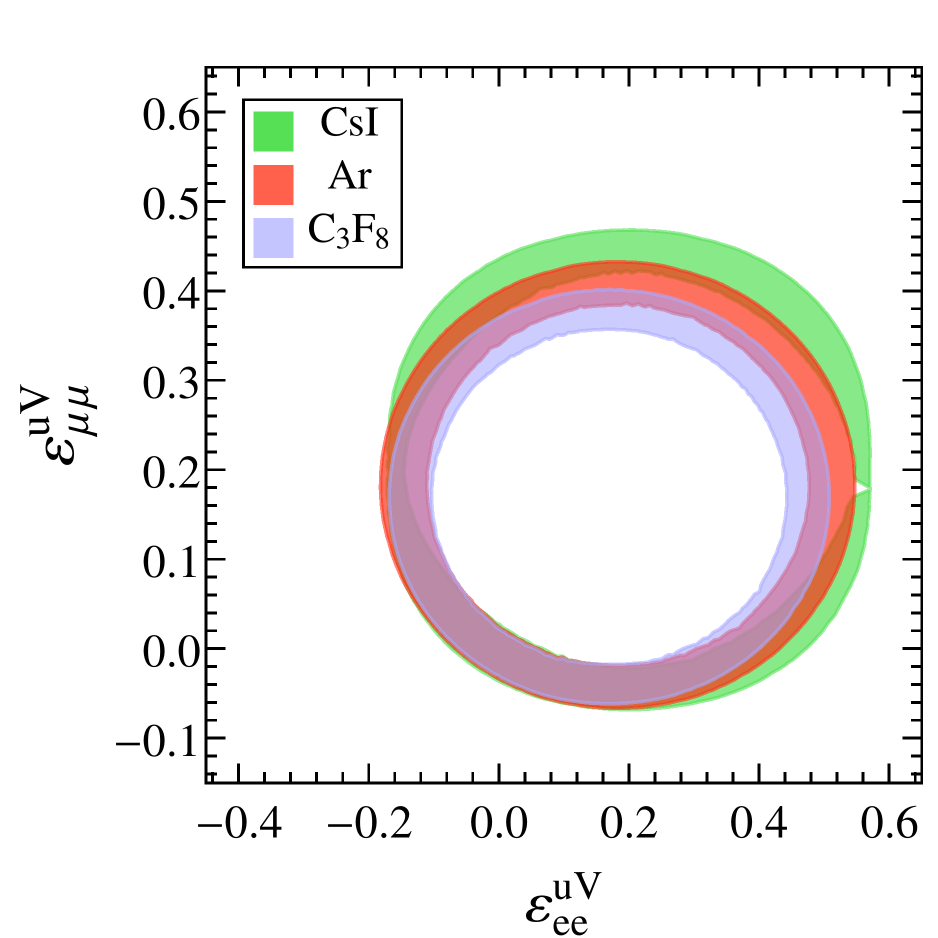}
\includegraphics[height=7.6cm, width=7.6cm]{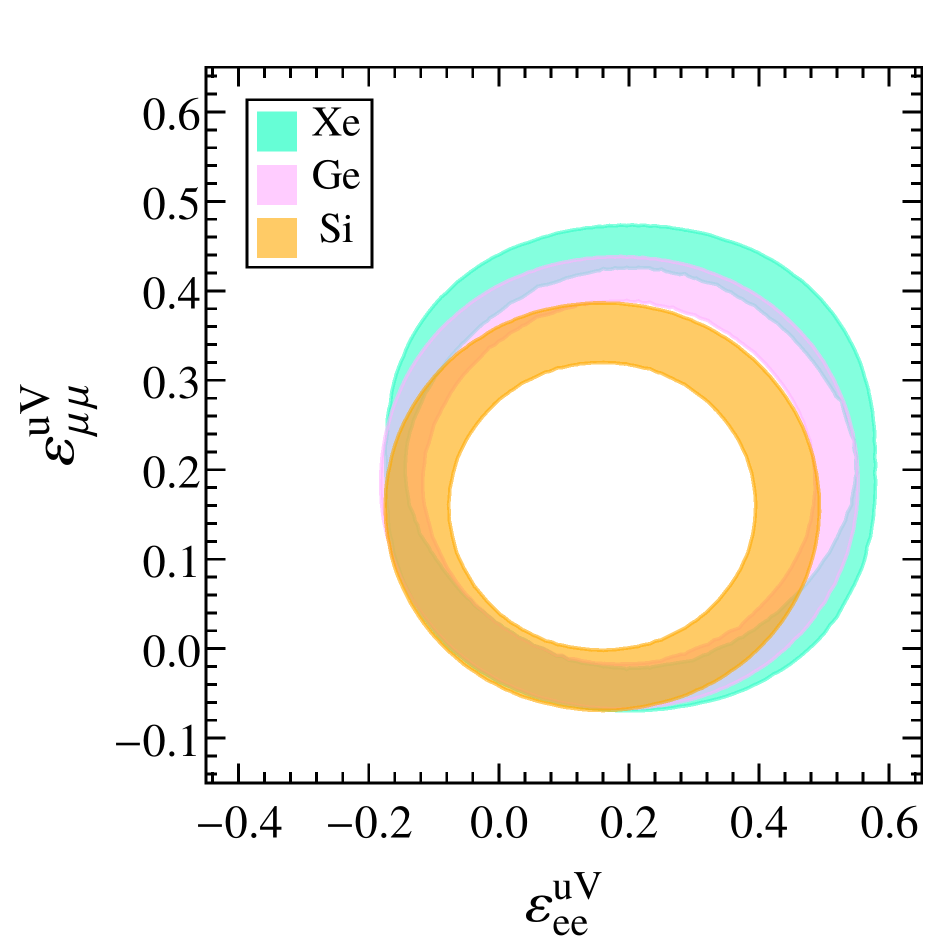}\\
\includegraphics[height=7.6cm, width=7.6cm]{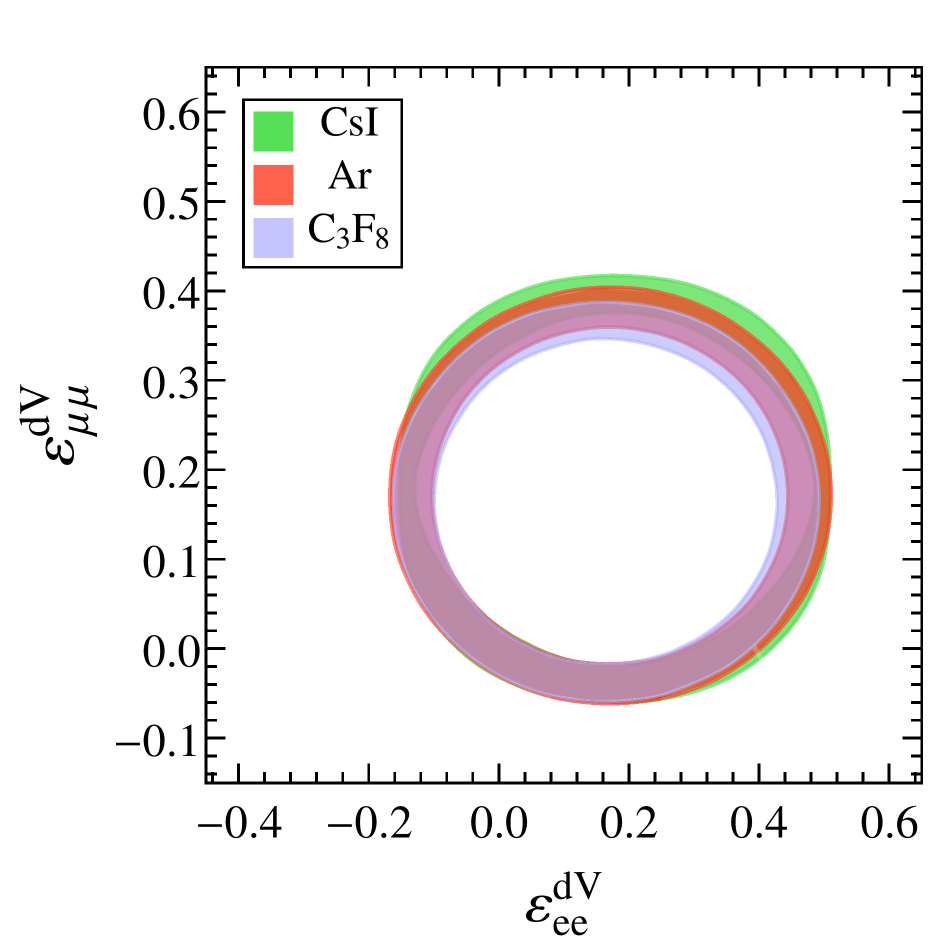}
\includegraphics[height=7.6cm, width=7.6cm]{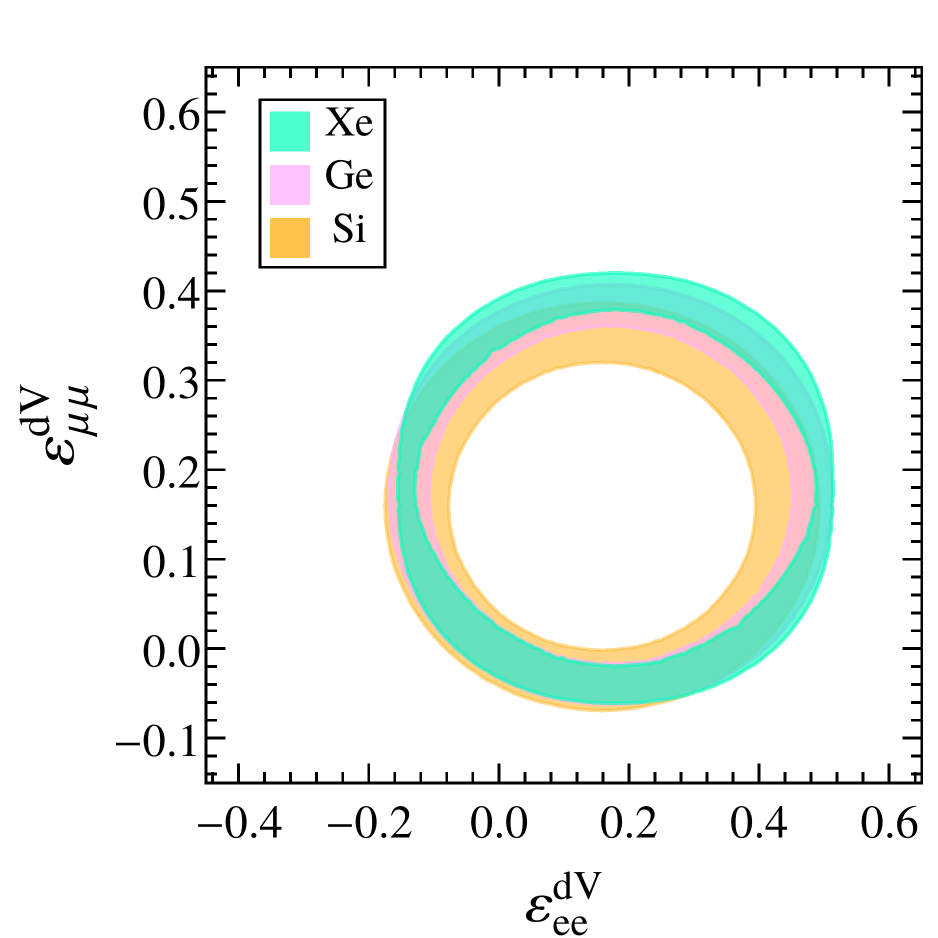}\\
\includegraphics[height=7.6cm, width=7.6cm]{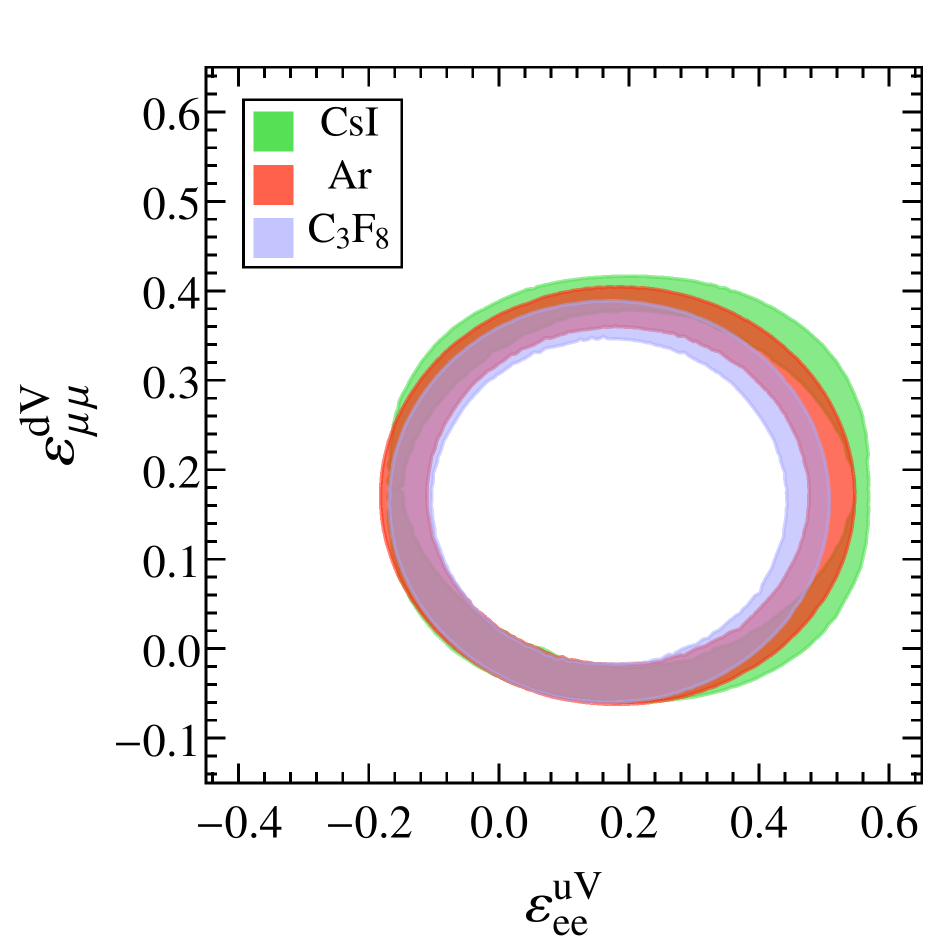}
\includegraphics[height=7.6cm, width=7.6cm]{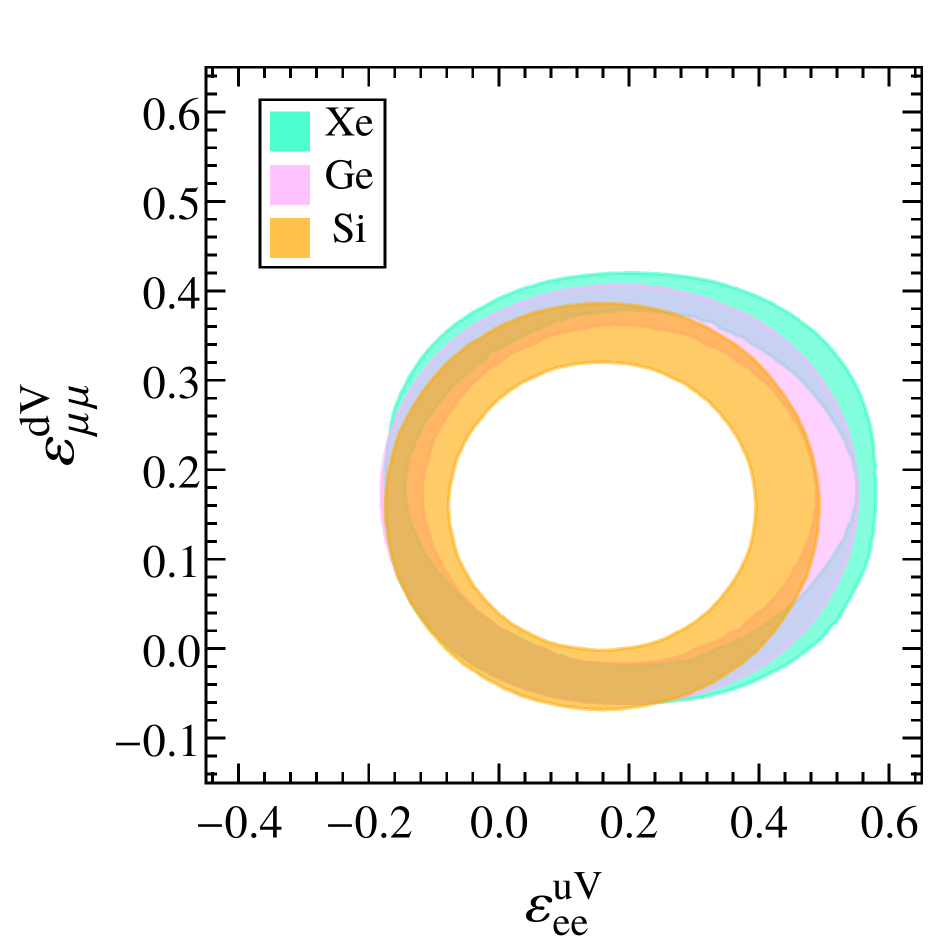}
\end{figure}
\begin{figure}[t]
\centering
\includegraphics[height=7.6cm, width=7.6cm]{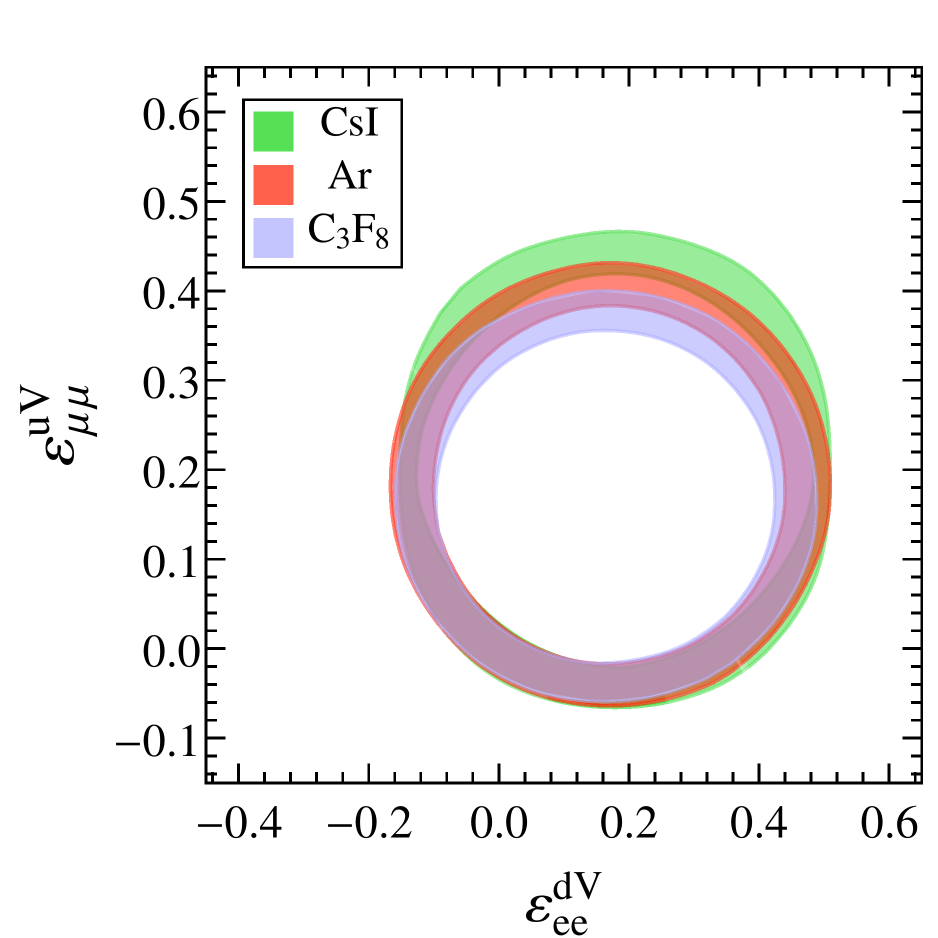}
\includegraphics[height=7.6cm, width=7.6cm]{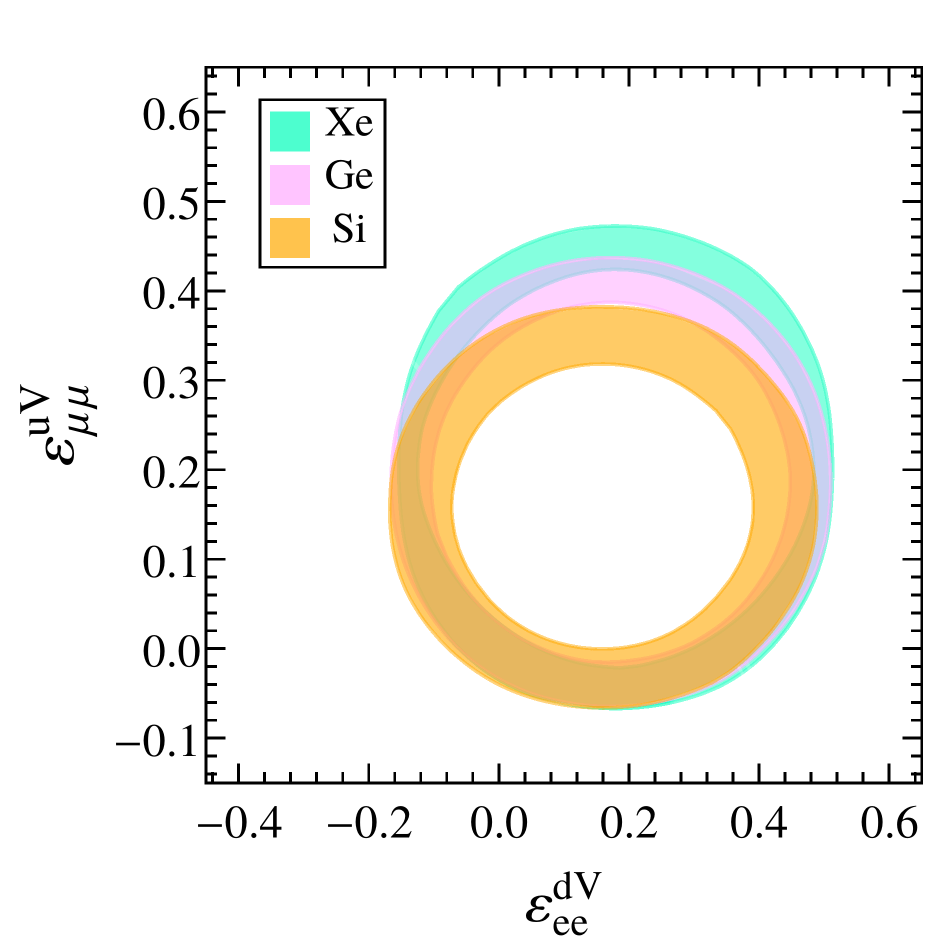}
\caption{Expected 90\% C. L. sensitivities (with two d.o.f., i.e. $\Delta \chi^2 \leq 4.61$)
to different combinations of two diagonal NSI parameters for
the detectors described in
Table~\ref{detectors_table1}. We assume 3 years
of data taking at the ESS.
Each color is associated with a specific
detector: CsI (green), Ar (red), Xe (cyan), Ge (magenta), Si (orange)
and C$_3$F$_8$ (blue). The plots in the first, second, third and
fourth rows correspond to the combinations
$(\varepsilon^{uV}_{ee},\, \varepsilon^{uV}_{\mu\mu})$,
$(\varepsilon^{dV}_{ee},\, \varepsilon^{dV}_{\mu\mu})$,
$(\varepsilon^{uV}_{ee},\, \varepsilon^{dV}_{\mu\mu})$ and
$(\varepsilon^{dV}_{ee},\, \varepsilon^{uV}_{\mu\mu})$, respectively.}
\label{fig:contour1}
\end{figure}

Let us first consider the individual detector sensitivities. In Fig.~\ref{fig:contour1}, we choose the two independent NSI parameters to be $\varepsilon^{qV}_{ee}$ and $\varepsilon^{q'V}_{\mu\mu}$, with $q, q' \in \{ u, d\}$, and we show the expected regions allowed by the different detectors at 90\% C.L. in the corresponding parameter space. Each row is associated with a different combination:
from top to bottom, $(\varepsilon^{uV}_{ee},\, \varepsilon^{uV}_{\mu\mu})$, $(\varepsilon^{dV}_{ee},\, \varepsilon^{dV}_{\mu\mu})$, $(\varepsilon^{uV}_{ee},\, \varepsilon^{dV}_{\mu\mu})$ and $(\varepsilon^{dV}_{ee},\, \varepsilon^{uV}_{\mu\mu})$. The green, red and blue regions in the left panels correspond to the CsI, Ar and $\rm C_3F_8$ detectors, while the cyan, magenta, and orange regions in the right panels correspond to the Xe, Ge, and Si detectors, respectively.
Notice that these colored regions share a characteristic elliptical (or approximatelly elliptical) shape. This is  easily  explained for single-target bubble chambers, like the Ar detector of Table~\ref{detectors_table1}, for which the analysis uses only the total number of events.  Assuming e.g. $\varepsilon^{uV}_{ee} \neq 0$ and $\varepsilon^{uV}_{\mu \mu} \neq 0,$ the predicted total number of signal events can be expressed as
\begin{equation}
N_{th} = C_e \left ( 2Z + N \right )^2 \left ( \varepsilon^{uV}_{ee} + \frac{Zg_{V}^p + Ng_{V}^n}{2Z + N} \right )^2 
+ C_\mu \left ( 2Z + N \right )^2 \left ( \varepsilon^{uV}_{\mu\mu} + \frac{Zg_{V}^p + Ng_{V}^n}{2Z + N} \right )^2 
,
\label{eq:ellipse}
\end{equation}
where $C_e$ ($C_\mu$) is the constant that results from performing the integral over $E_{\nu_e}$ ($E_{\nu_\mu}$ and $E_{\bar \nu_\mu}$) in Eq.~(\ref{eq:events}), after factorizing the weak nuclear charge (given by Eq.~(\ref{eq:weak_charge_NSIs})) out of the CE$\nu$NS cross-section. For a fixed number of events $N_{th}$, Eq. (\ref{eq:ellipse}) is the equation of an ellipse with minor and major semi-axes given by $\sqrt{N_{th}/C_e \left( 2Z + N \right)^2}$ and $\sqrt{N_{th}/C_\mu \left( 2Z + N \right)^2}$, respectively.
Values of $\varepsilon^{uV}_{ee}$ and $\varepsilon^{uV}_{\mu \mu}$ lying on the same ellipse reproduce the same  number of events $N_{th}$, hence the same value of $\chi^2$ in Eq. (\ref{eq:analysis}): they are degenerate from the point of view of CE$\nu$NS measurements. The expected 90\% C.L. sensitivity region associated with the Ar detector in the first row of Fig.~\ref{fig:contour1} is therefore the collection of all ellipses satisfying $\Delta\chi^2 \leq 4.61$.
The same arguments apply to the other three combinations of the $\varepsilon^{qV}_{ee}$ and $\varepsilon^{qV} _{\mu \mu}$ parameters, and explain the elliptical shape of the Ar regions in Fig.~\ref{fig:contour1}.

The above discussion can be generalized to bubble chamber detectors with two nuclei, like $\textrm{C}_3\textrm{F}_8$, where it is also possible to write the predicted number of events as the equation of an ellipse, with more complicated expressions for the axes. For recoil energy-sensitive detectors (CsI, Xe, Si and Ge), the analysis is done bin by bin, and the ellipses defined by Eq.~(\ref{eq:ellipse}) do not necessarily explain accurately the shape of the 90\% C.L. sensitivity regions displayed in Fig.~\ref{fig:contour1}. Indeed, while two points on one of these ellipses correspond to the same total number of events, they may not predict the same recoil energy distribution, hence their $\chi^2$ value may slightly differ. This explains, for instance, why the sensitivity regions for CsI and Xe are squeezed at some points.

\begin{figure}[t]
\centering
\includegraphics[height=7.6cm, width=7.6cm]{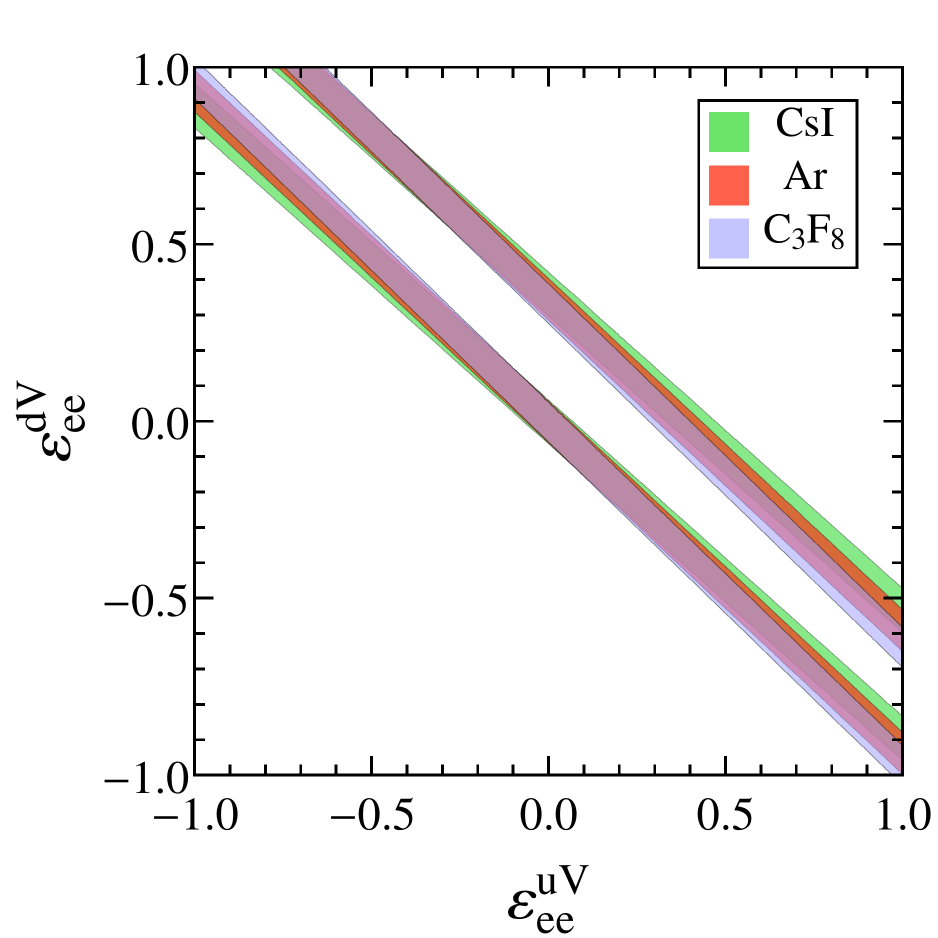}
\includegraphics[height=7.6cm, width=7.6cm]{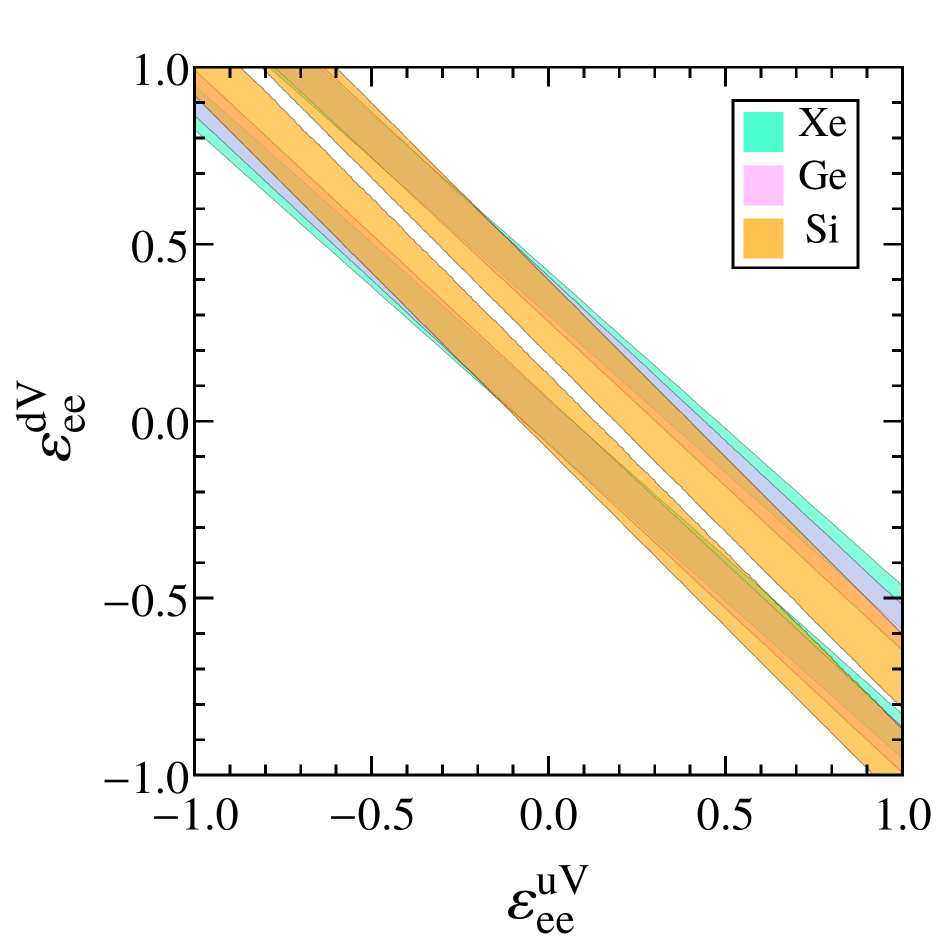}\\
\includegraphics[height=7.6cm, width=7.6cm]{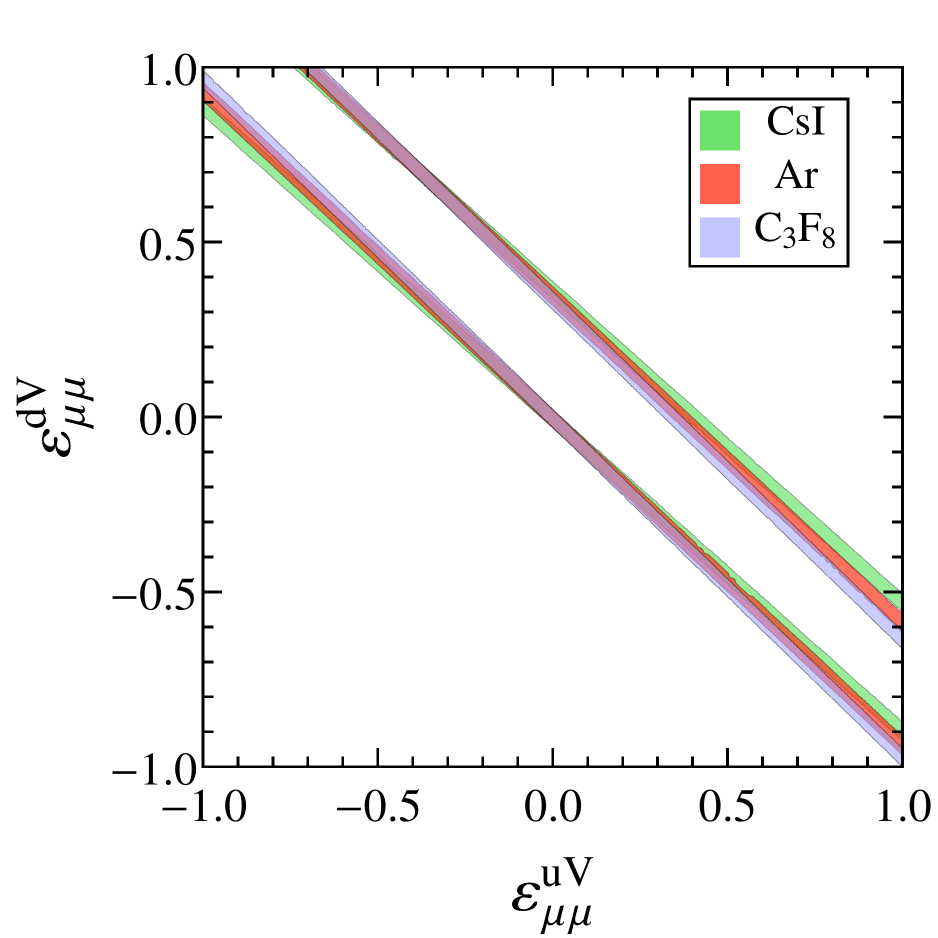}
\includegraphics[height=7.6cm, width=7.6cm]{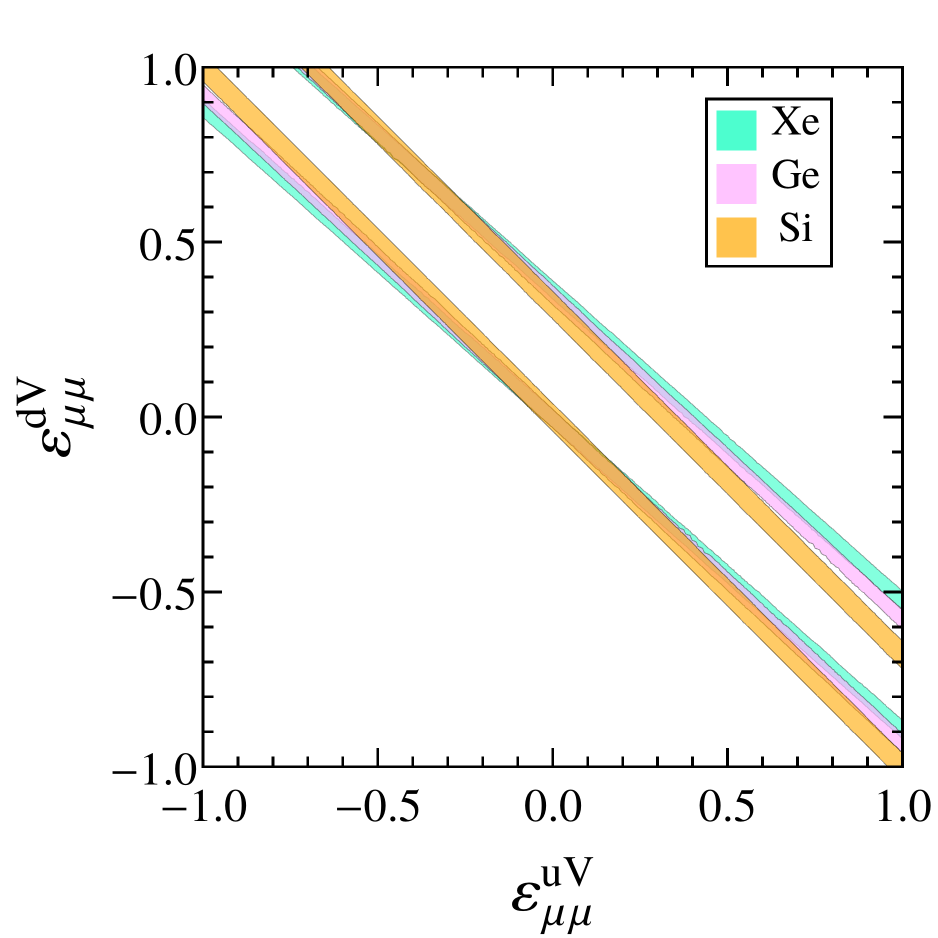}
\caption{Expected 90\% C. L. sensitivities (with two d.o.f., i.e. $\Delta \chi^2 \leq 4.61$)
to different combinations of two diagonal NSI parameters for
the detectors described in
Table~\ref{detectors_table1}. We assume 3 years
of data taking at the ESS.
The color code is the same as in Fig. \ref{fig:contour1}. The top panels correspond to the combinations $(\varepsilon^{uV}_{ee},\, \varepsilon^{dV}_{ee})$, and the bottom panels to $(\varepsilon^{uV}_{\mu\mu},\, \varepsilon^{dV}_{\mu\mu})$.}
\label{fig:contour2}
\end{figure}

In Fig.~\ref{fig:contour2}, we show the expected 90\% C.L. sensitivities for
the remaining two possible combinations of diagonal NSI parameters,
namely
$\left(\varepsilon^{uV}_{\alpha \alpha},\, \varepsilon^{dV}_{\alpha\alpha} \right)$
with $\alpha = e$ or $\mu$. These regions are composed of two parallel
linear bands~\cite{Scholberg:2005qs}, a fact that can be explained
again from the expressions of the weak nuclear charge and
differential CE$\nu$NS event rate in Eqs.~(\ref{eq:weak_charge_NSIs})
and~(\ref{eq:events}), respectively. Indeed, assuming for
instance the two nonvanishing NSI parameters to be
$\varepsilon^{uV}_{ee}$ and $\varepsilon^{dV} _{ee}$, we can write the
predicted total number of signal events as (for a single-nucleus
detector)
\begin{equation}
N_{th} = \left [ Zg_V^{p} + Ng_V^n + \left ( 2Z+N \right )\varepsilon^{uV}_{ee} + \left ( Z+2N \right )\varepsilon^{dV} _{ee}\right ]^2C_e +  \left [ Zg_V^{p} + Ng_V^n \right ]^2C_\mu\; ,
\label{eq:line1}
\end{equation}
where $C_{e}$ and $C_\mu$ are the same constants that appear in Eq. (\ref{eq:ellipse}). For fixed $N_{th}$, Eq. (\ref{eq:line1}) describes two straight lines in the $\left(\varepsilon^{uV}_{ee},\, \varepsilon^{dV}_{ee} \right)$ plane, with a common slope $m$ given by~\cite{Barranco:2005yy}
\begin{equation}
m = -\frac{2Z+N}{Z+2N}\; 
\label{slope}
\end{equation}
(obviously, a similar conclusion holds if the two nonvanishing NSI parameters are chosen to be $\varepsilon^{uV}_{\mu\mu}$ and  $\varepsilon^{dV} _{\mu\mu}$ rather than $\varepsilon^{uV}_{ee}$ and  $\varepsilon^{dV} _{ee}$).
For recoil energy-sensitive detectors, there is an equation analogous to Eq.~(\ref{eq:line1}) for each bin $i$, with $N_{th}$, $C_{e}$ and $C_\mu$ replaced by bin-dependent quantities $N^i_{th}$, $C^i_{e}$ and $C^i_\mu$, but the slope, still given by Eq.~(\ref{slope}), is the same for all bins\footnote{The situation is more subtle for detectors made of two nuclei, but in practice the slopes associated with each nucleus are very close, both for CsI and for C$_3$F$_8$.}. Thus, irrespective of the detection technique, each expected 90\% C.L. allowed region in Fig.~\ref{fig:contour2} is the collection of all line intervals with slope $m$ satisfying $\Delta\chi^2 \leq 4.61$ (where both $m$ and the function $\Delta\chi^2 = \chi^2(\varepsilon^{uV}_{ee}, \varepsilon^{dV}_{ee})$ depend on the detector). This explains why these regions are made of two parallel bands.

\begin{figure}[t]
\centering
\includegraphics[height=7.6cm, width=7.6cm]{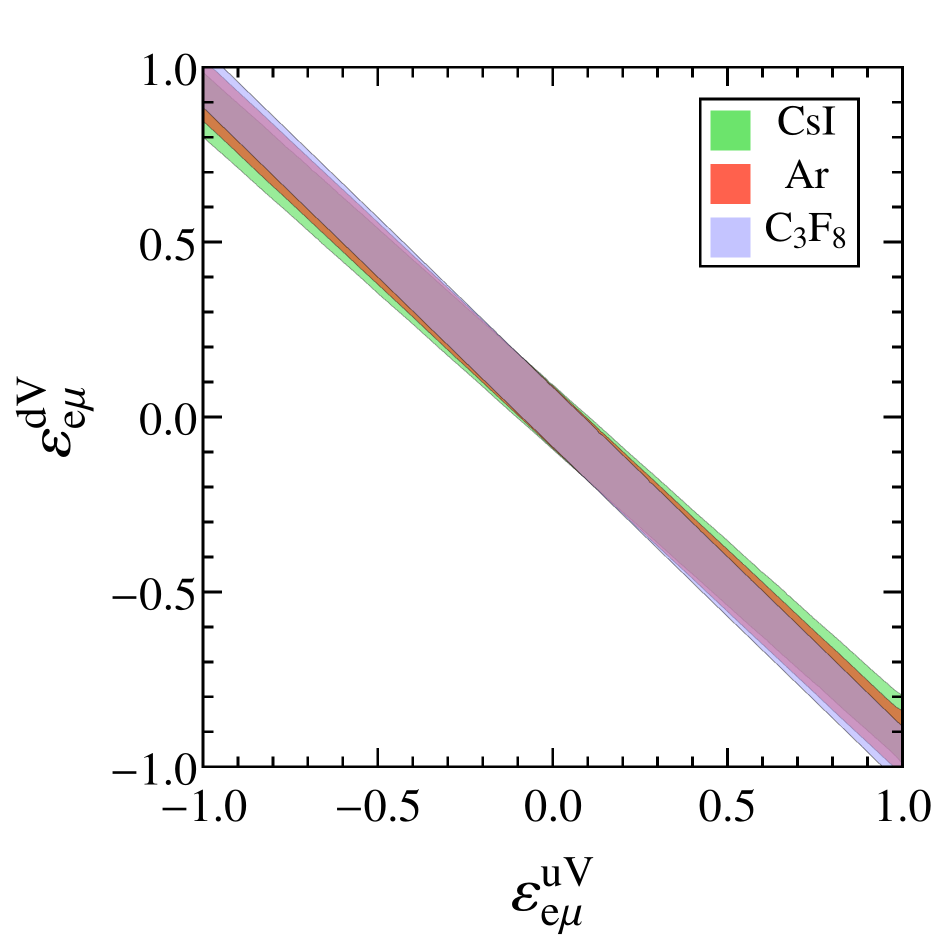}
\includegraphics[height=7.6cm, width=7.6cm]{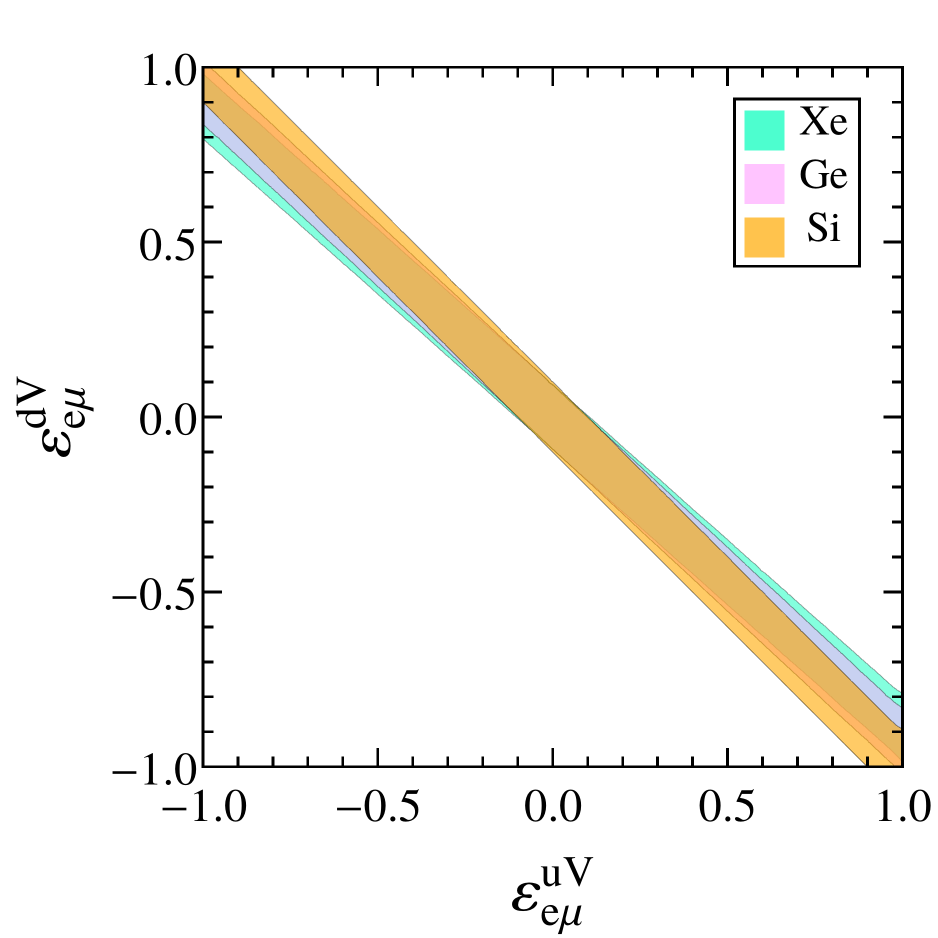}\\
\includegraphics[height=7.6cm, width=7.6cm]{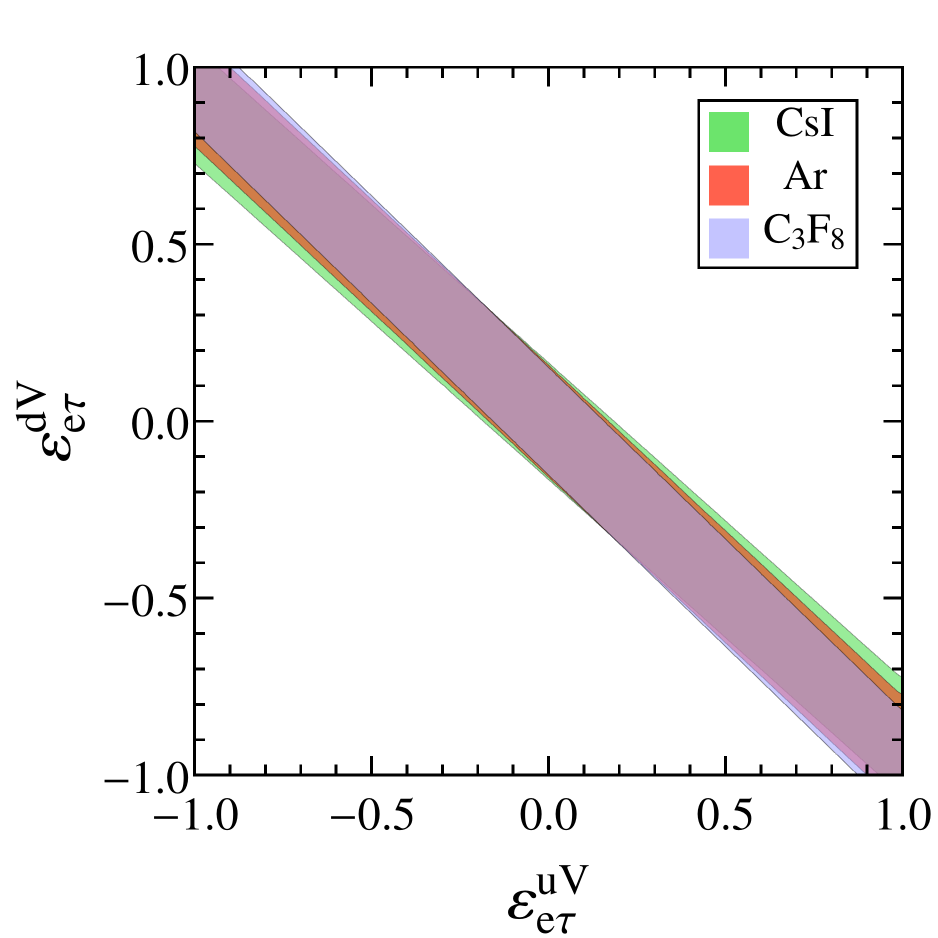}
\includegraphics[height=7.6cm, width=7.6cm]{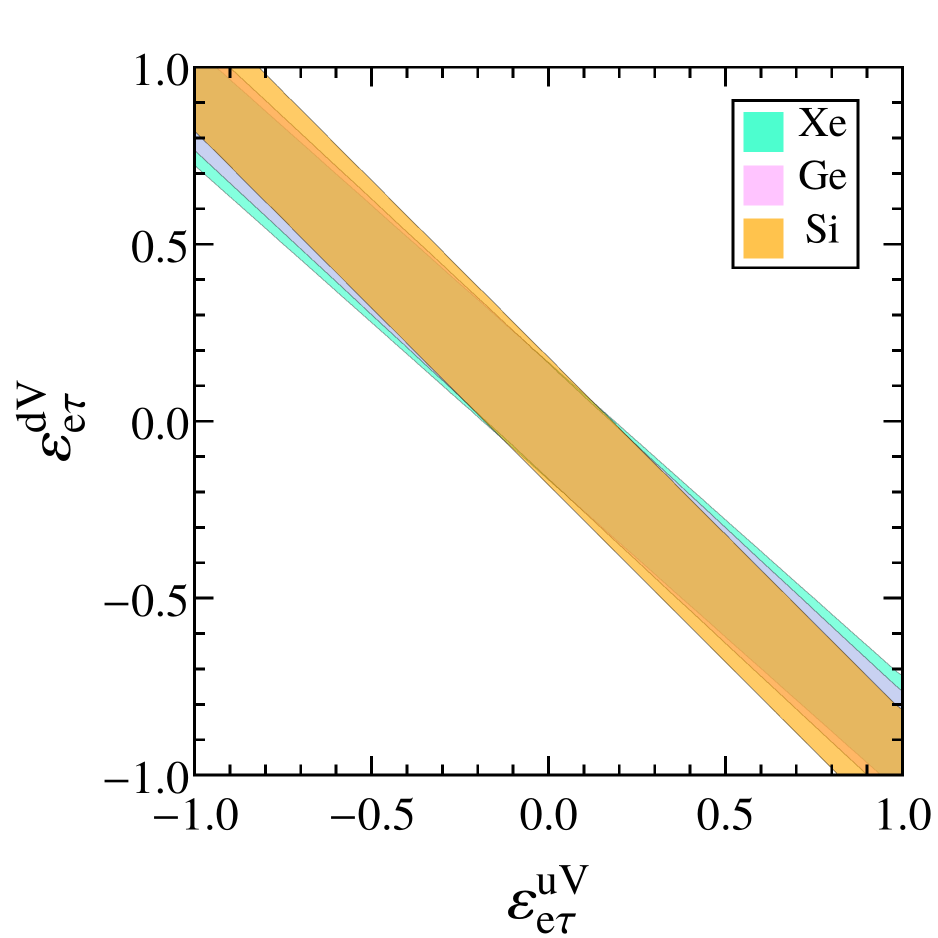}\\
\includegraphics[height=7.6cm, width=7.6cm]{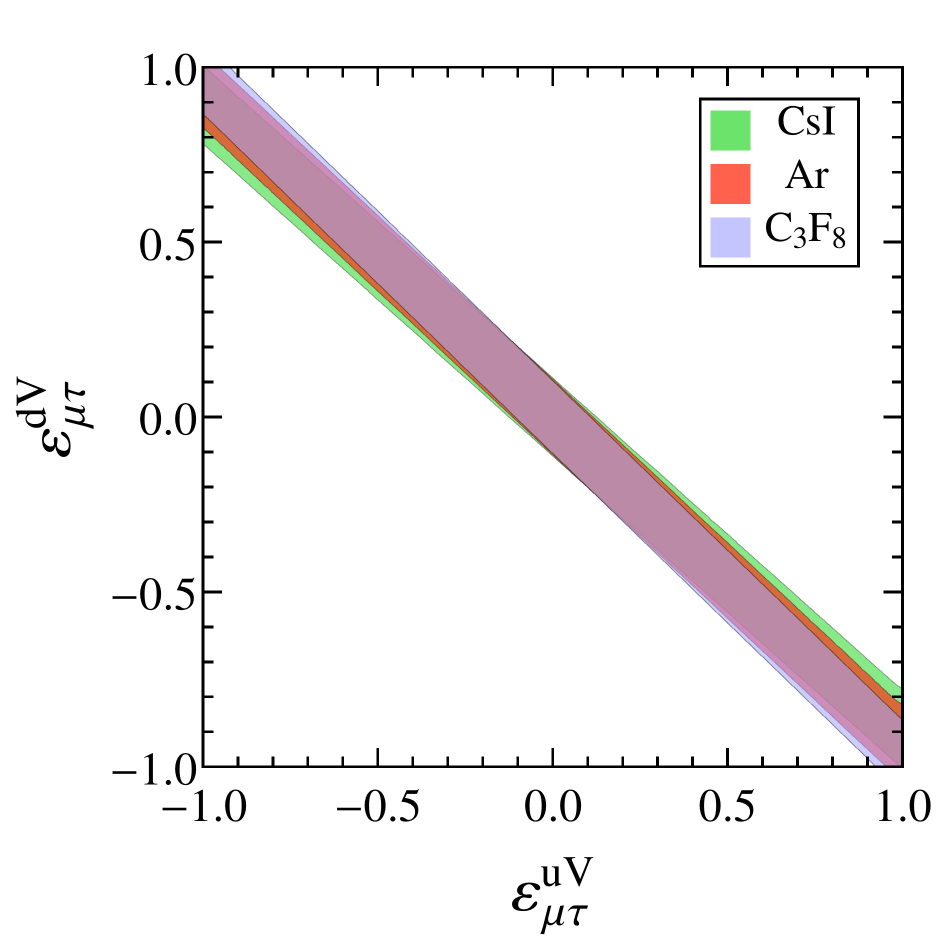}
\includegraphics[height=7.6cm, width=7.6cm]{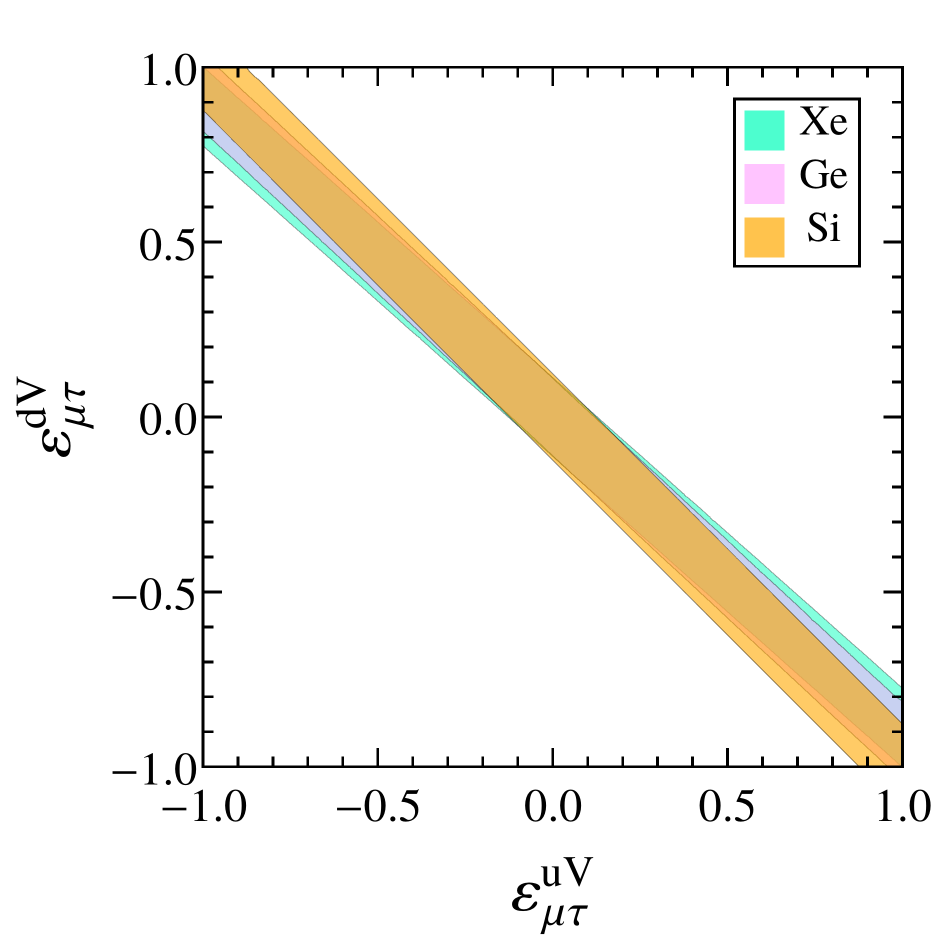}
\caption{Expected 90\% C. L. sensitivities (with two d.o.f., i.e. $\Delta \chi^2 \leq 4.61$)
to different combinations of two diagonal NSI parameters for
the detectors described in
Table~\ref{detectors_table1}.  We assume 3 years
of data taking at the ESS.
The color code is the same as in Fig. \ref{fig:contour1}.
The top, central and bottom panels correspond to the combinations $(\varepsilon^{uV}_{e\mu},\, \varepsilon^{dV}_{e\mu})$, $(\varepsilon^{uV}_{e\tau},\, \varepsilon^{dV}_{e\tau})$ and $(\varepsilon^{uV}_{\mu\tau},\, \varepsilon^{dV}_{\mu\tau})$, respectively.}
\label{fig:contour3}
\end{figure}

Finally, in Fig.~\ref{fig:contour3}, we show the expected 90\% C.L. sensitivities for the combinations of off-diagonal NSI parameters
  $\left(\varepsilon^{uV}_{\alpha \beta},\, \varepsilon^{dV}_{\alpha\beta} \right)$, $\alpha \neq \beta$. In this case, the sensitivity regions are single linear bands, which can be understood from arguments similar to the ones used above. Indeed, assuming the two nonvanishing NSI parameters to be $\varepsilon^{uV}_{e \mu}$ and  $\varepsilon^{dV} _{e \mu}$, one can write the expected total number of signal events in the form (for a single-nucleus detector)
\begin{equation}
N_{th}\, =\, C \left\{ \left [ Zg_V^{p} + Ng_V^n \right ]^2 + \left [ \left ( 2Z+N \right )\varepsilon^{uV}_{e\mu} + \left ( Z+2N \right )\varepsilon^{dV}_{e\mu} \right ]^2 \right\} ,
\label{eq:line2}
\end{equation}
where $C$ is the constant that results from performing the integral over all three neutrino fluxes in Eq.~(\ref{eq:events}) after factorizing out the weak nuclear charge. For a fixed number of events $N_{th}$, Eq.~(\ref{eq:line2}) describes two straight lines in the $\left(\varepsilon^{uV}_{e\mu},\, \varepsilon^{dV}_{e\mu} \right)$ plane, symmetric under a reflection through the origin and with a slope $m$ again given by Eq.~(\ref{slope}). For $N_{th} = N^{SM}_{th} = C (Zg_V^{p} + Ng_V^n)^2$, the two lines merge into a single one crossing the origin. Similar conclusions hold if the two nonvanishing NSI parameters are chosen to be ($\varepsilon^{uV}_{e\tau}$, $\varepsilon^{dV} _{e\tau}$) or ($\varepsilon^{uV}_{\mu\tau}$, $\varepsilon^{dV} _{\mu\tau}$). Omitting detailed explanations analogous to the ones given for Fig.~\ref{fig:contour2}, this explains why the expected 90\% C.L. sensitivity regions in Fig.~\ref{fig:contour3} consist of a single linear band of slope $m$ containing the origin.

 \begin{figure}[t]
\centering
\includegraphics[height=5.4cm, width=5.4cm]{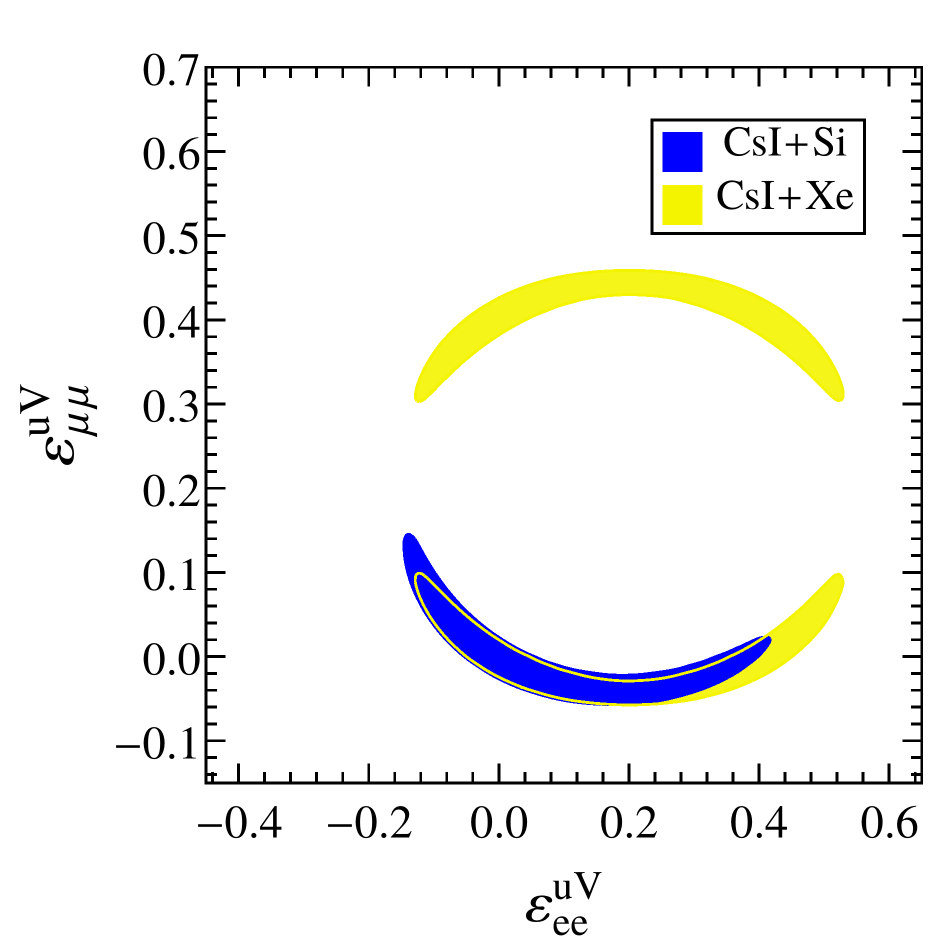}
\includegraphics[height=5.4cm, width=5.4cm]{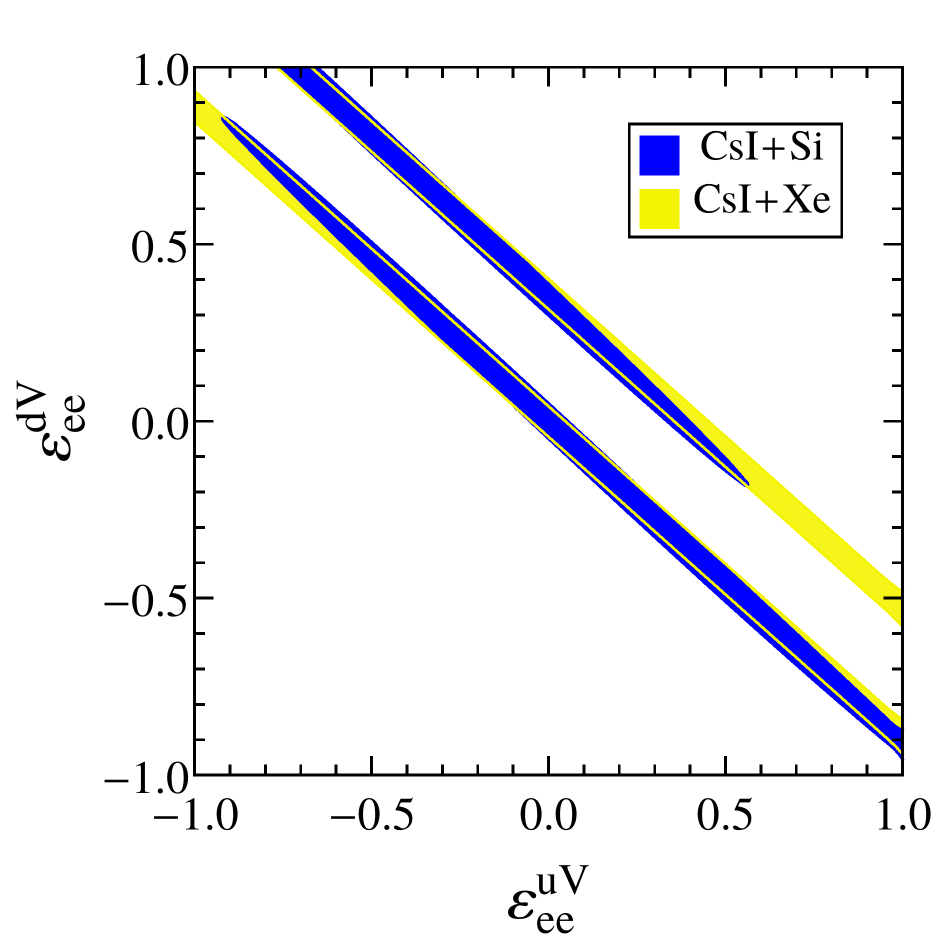}
\includegraphics[height=5.4cm, width=5.4cm]{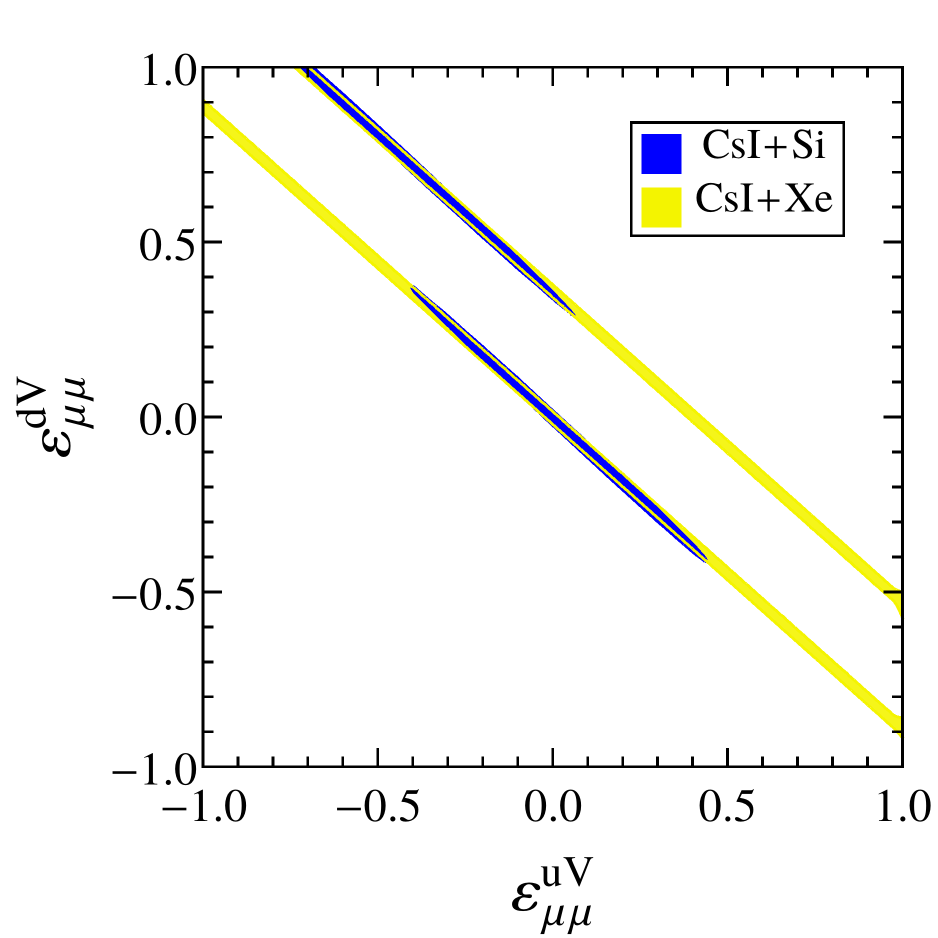}
\caption{Expected 90\% C. L. sensitivities (with two d.o.f., i.e. $\Delta \chi^2 \leq 4.61$), in the planes defined by different sets of two diagonal NSI parameters, allowed by the combinations of detectors $\rm CsI + Xe$ (yellow) and $\rm CsI + Si$ (blue), assuming 3 years of data taking at the ESS for each detector. The left, central and right panels correspond to the parameters $(\varepsilon^{uV}_{ee},\, \varepsilon^{uV}_{\mu\mu})$, $(\varepsilon^{uV}_{ee},\, \varepsilon^{dV}_{ee})$ and $(\varepsilon^{uV}_{\mu\mu},\, \varepsilon^{dV}_{\mu\mu})$, respectively.}
\label{fig:contour4}
\end{figure}

As can be seen from Figs.~\ref{fig:contour1}, \ref{fig:contour2}
and~\ref{fig:contour3}, degeneracies between different NSI parameters
appear whenever one tries to constrain more than one parameter at a
time from a single CE$\nu$NS measurement experiment. We now show how
these degeneracies can be partially lifted when combining the results
of two different detectors. Given the large number of possible
combinations of two detectors, we restrict our study to a few suitably
chosen cases. Already from inspecting how the different sensitivity
regions in Figs.~\ref{fig:contour1} to~\ref{fig:contour3} overlap, one
can identify some more favorable combinations. In particular,
detectors with a different proton to neutron 
ratio~\cite{Barranco:2005yy} help to reduce the degeneracies in
Figs.~\ref{fig:contour2} and~\ref{fig:contour3}, as the associated
bands are characterized by different slopes (indeed, $m$ can be
written as $m = - (2r+1)(r+2)$, where $r = Z/N$). As an illustration,
we show in Fig.~\ref{fig:contour4} the expected 90\% C.L. sensitivities in
the parameter spaces
$\left(\varepsilon^{uV}_{ee},\, \varepsilon^{uV}_{\mu\mu} \right)$,
$\left(\varepsilon^{uV}_{ee},\, \varepsilon^{dV}_{ee}\right)$ and
$\left(\varepsilon^{uV}_{\mu\mu},\, \varepsilon^{dV}_{\mu\mu} \right)$
for the detector combinations $\rm CsI+Si$, and $\rm CsI+Xe$. As can
be seen, the most efficient combination is the one involving two
target materials with a very different proton to
neutron ratio ($r = 0.71$
[0.72] for Cs [I] versus $r = 1$ for Si, while Xe has $r = 0.69$, very
close to the Cs and I nuclei).
Moreover, by comparing the blue area on the left plot of Fig.~\ref{fig:contour4}
with Fig.~14 of Ref.~\cite{Baxter:2019mcx}, one can see that combining the data of two detectors
with different enough characteristics, such as CsI and Si, is much more efficient in breaking
degeneracies than filling the same gas TPC detector alternatively with Xe and Ar,
which is the option studied in Ref.~\cite{Baxter:2019mcx}. To make the comparison more meaningful,
we checked that this conclusion still holds if the CsI and Si detectors run only for 1.5 years each,
as for the gas TPC detector filled with either Xe or Ar.

\begin{figure}[t!]
\centering
\includegraphics[height=5.4cm, width=5.4cm]{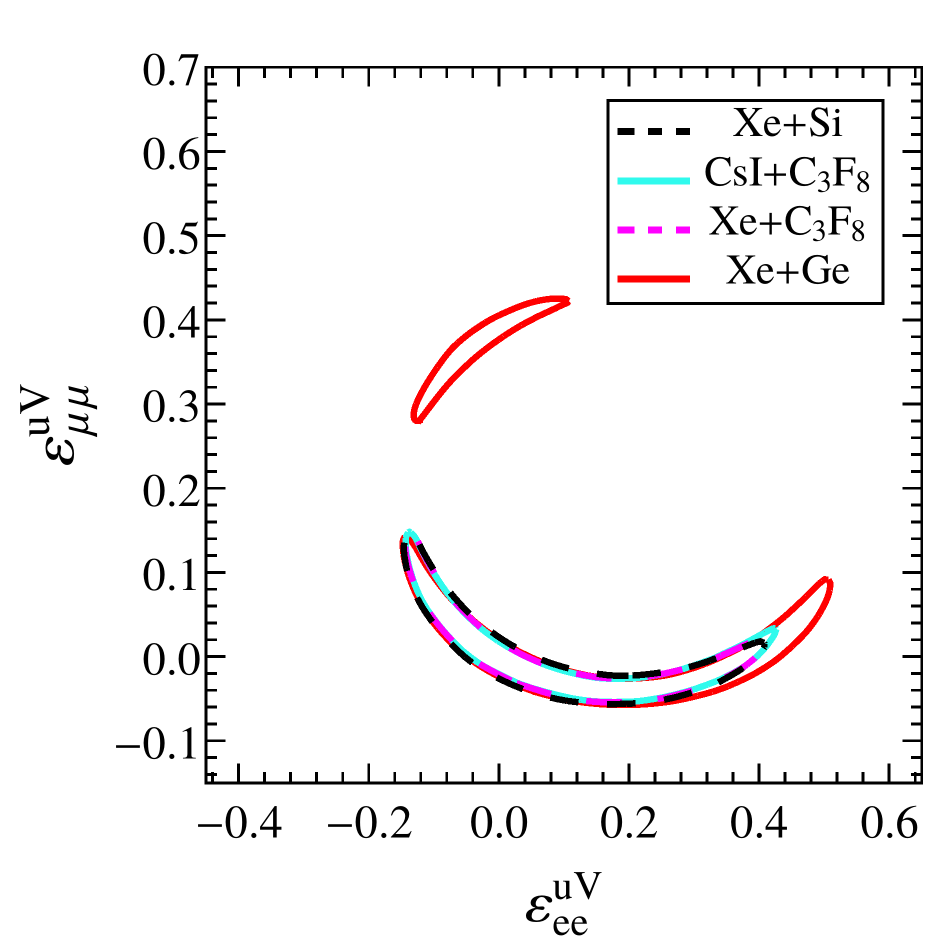}
\includegraphics[height=5.4cm, width=5.4cm]{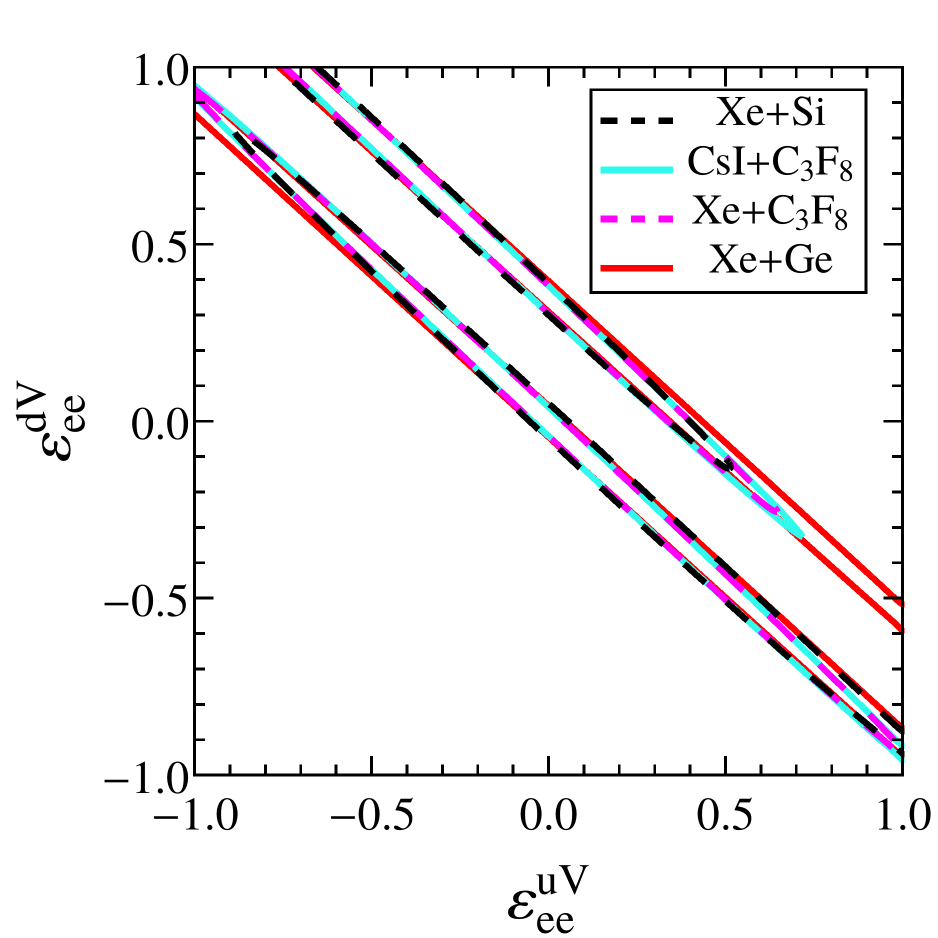}
\includegraphics[height=5.4cm, width=5.4cm]{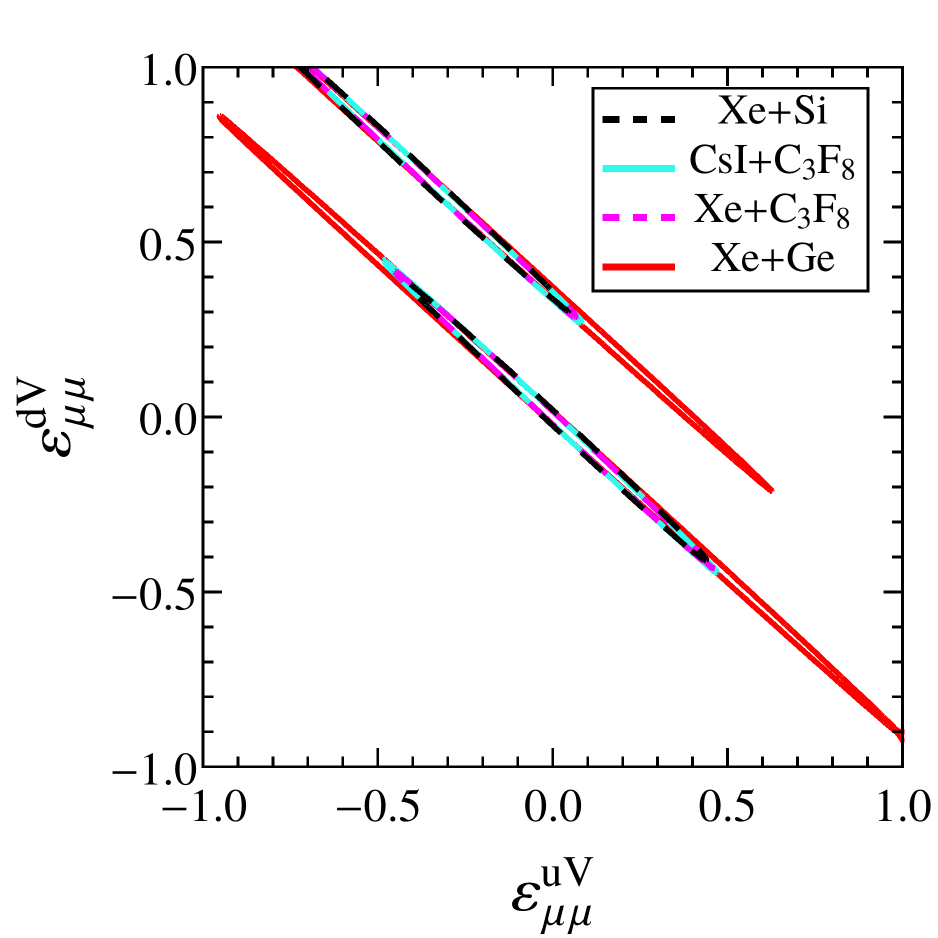}
\caption{Same as Fig.~\ref{fig:contour4} for different combinations of two detectors: $\rm Xe + Si$ (dashed black), $\rm CsI + C_3F_8$ (solid cyan), $\rm Xe + C_3F_8$ (dashed magenta) and $\rm Xe + Ge$ (solid red).}
\label{fig:contour5}
\end{figure}

For comparison, we show in Fig.~\ref{fig:contour5} a few more combinations of two detectors, again chosen among the most suitable ones to reduce degeneracies between two different NSI parameters. 
Among these combinations, $\rm Xe + Si$ is comparable to $\rm CsI + Si$, while 
$\rm CsI + C_3F_8$ and $\rm Xe + C_3F_8$ are almost as efficient. On the other hand, $\rm Xe + Ge$ is significantly less sensitive, but still better than $\rm CsI + Xe$. These results are expected from the previous discussion regarding the impact of the proton to neutron ratio of the targets: the combinations that have very different ratios are more suitable to break degeneracies between NSI parameters. 

\begin{figure}[t!]
\centering
\includegraphics[height=5.4cm, width=5.4cm]{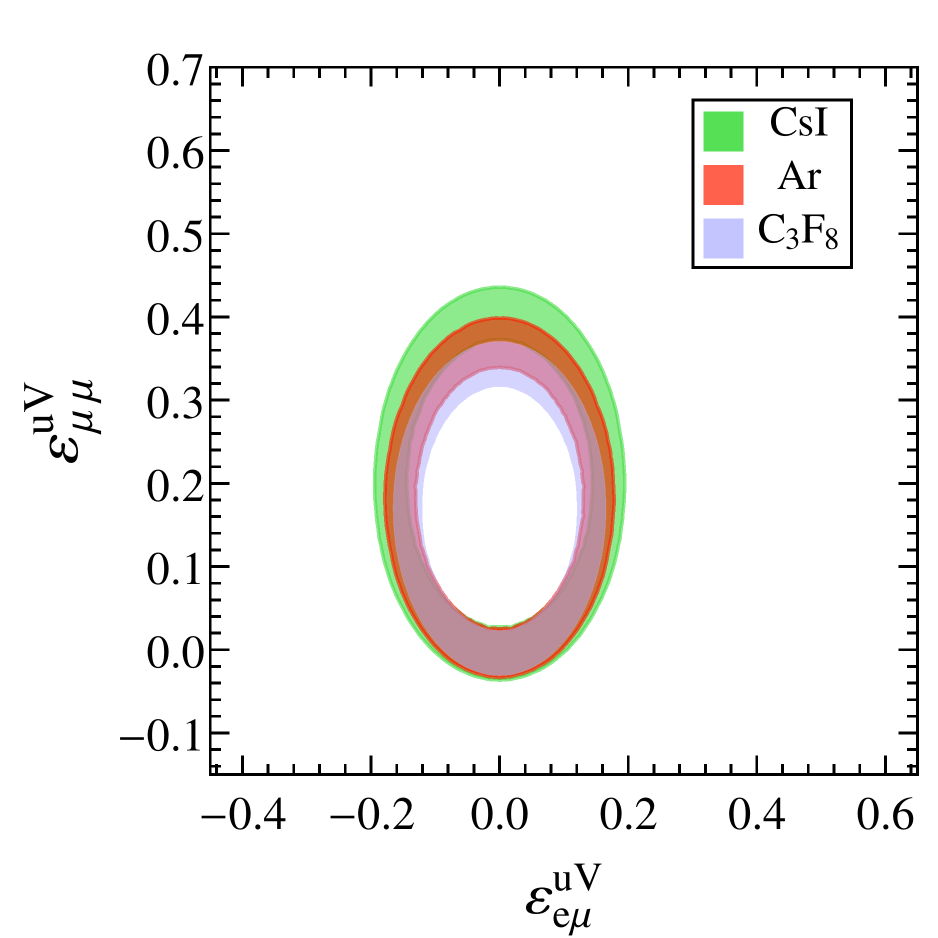}
\includegraphics[height=5.4cm, width=5.4cm]{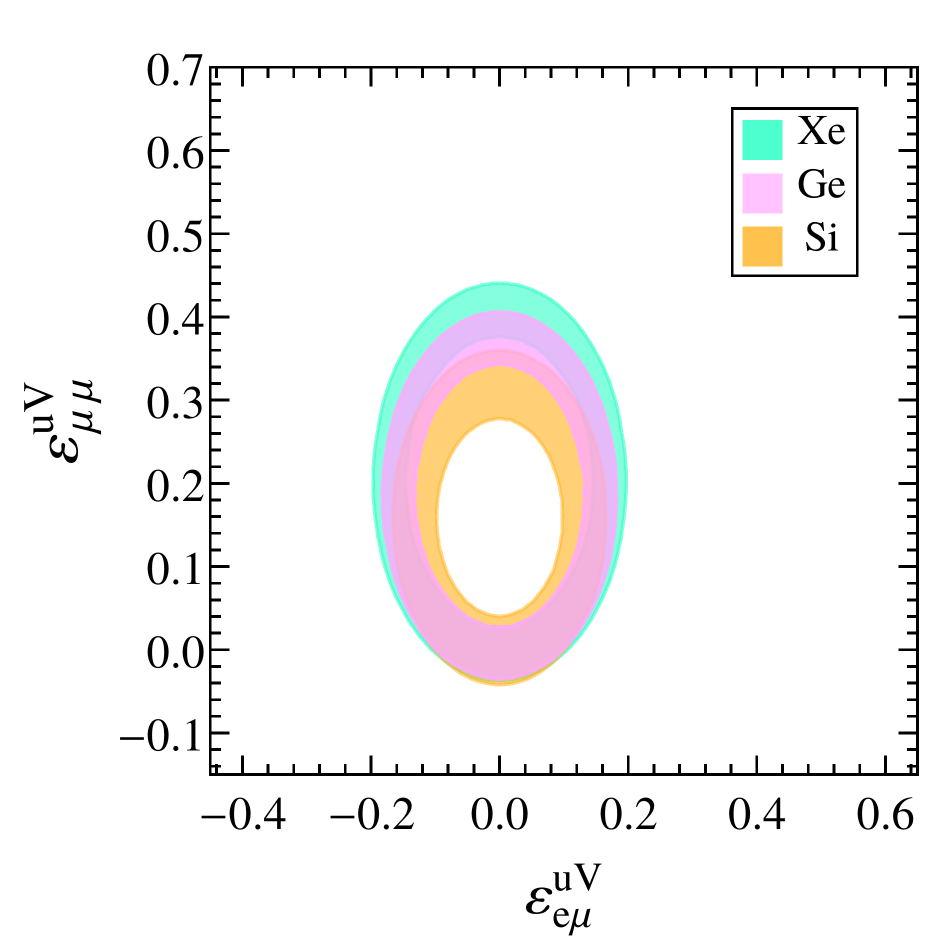}
\includegraphics[height=5.4cm, width=5.4cm]{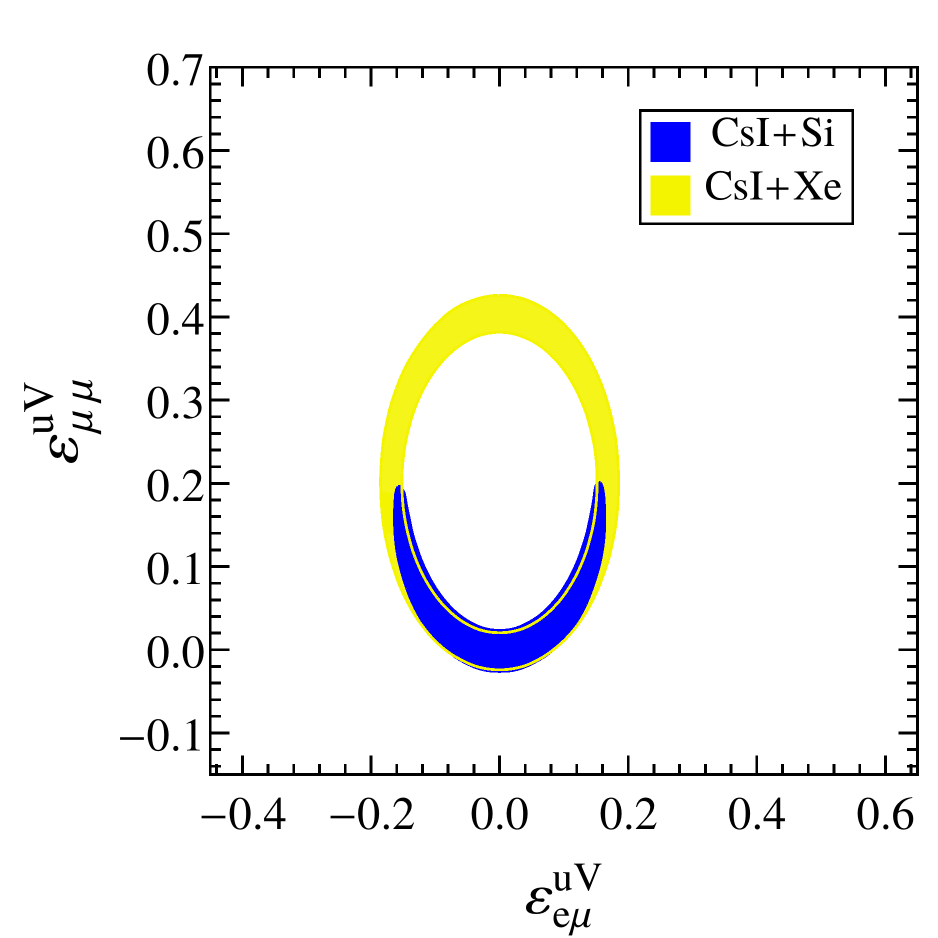}
\caption{Expected 90\% C. L. sensitivities (with two d.o.f., i.e. $\Delta \chi^2 \leq 4.61$) in the ($\varepsilon^{uV}_{e\mu}$, $\varepsilon^{uV}_{\mu\mu}$) plane corresponding to different detectors or combination of detectors, assuming 3 years of data taking at the ESS for each detector.  Left: CsI (green), Ar (red) and C$_3$F$_8$ (blue). Middle: Xe (cyan), Ge (magenta) and Si (orange). Right: combinations of detectors $\rm CsI + Xe$ (yellow) and $\rm CsI + Si$ (blue).}
\label{fig:contour6}
\end{figure}

Finally, we briefly comment on degeneracies between off-diagonal NSI parameters.
As can be seen from Fig.~\ref{fig:contour3}, the sensitivity regions for the pairs of NSI parameters ($\varepsilon^{uV}_{\alpha \beta}$, $\varepsilon^{dV}_{\alpha \beta}$), $\alpha \neq \beta$, are just a little bit tilted with respect to each other and overlap to a large extent\footnote{This large overlap is due to the fact that the contribution of off-diagonal NSI parameters to the CE$\nu$NS cross section arises at order $\varepsilon^2$, as can be seen from Eq.~(\ref{eq:line2}). By contrast, the contribution of diagonal NSI parameters is of order $\varepsilon$ due to the interference with the SM contribution (see Eq.~(\ref{eq:line1})).}. Therefore, combining the data from two different detectors does not lead to a significant reduction of degeneracies between these parameters.
An even stronger conclusion holds for pairs of off-diagonal NSI parameters with different flavour indices, such as ($\varepsilon^{uV}_{e \mu}$, $\varepsilon^{uV}_{e \tau}$), whose contributions to the CE$\nu$NS cross section do not interfere: in this case, the only benefit of combining the data from two different detectors is the increase in statistics.
However, when a diagonal and an off-diagonal NSI parameters are assumed to be simultaneously nonvanishing, combining two suitably chosen detectors does lead to a partial breaking of degeneracies, as can be seen from the plots in Fig.~\ref{fig:contour6}.

%


\section{Conclusions}
\label{sec:conclusions}

In this work, we explored the physics potential of the European Spallation Source setup in constraining non-standard neutrino interactions using coherent elastic neutrino-nucleus scattering. More specifically, we studied the sensitivity that different detectors can achieve on NC-NSI parameters after 3 years of data taking, using the ESS neutrino beam as a source for high-statistics CE$\nu$NS measurements. These detectors, proposed in Ref.~\cite{Baxter:2019mcx}, use different detection technologies with various target materials (CsI, Xe, Si, Ge, Ar  and C$_3$F$_8$) and are characterized by different recoil energy thresholds. We first showed that these detectors could ameliorate the bounds on individual NSI parameters  extracted from the COHERENT data. Then we quantified the further improvement that can be achieved when two detectors operate simultaneously. We found that the detector combination $\rm CsI+Si$ has one of the best sensitivities to the individual NSI parameters, and that the improvement with respect to a single target material is particularly significant for the flavor-diagonal parameters $\varepsilon^{uV}_{ee}$, $\varepsilon^{dV}_{\mu\mu}$ and especially $\varepsilon^{uV}_{\mu\mu}$. The combined analysis of the results of two detectors is an even more powerful tool when two NSI parameters are assumed to be nonvanishing at a time. We considered different detector combinations and studied their ability to partially break the degeneracies between two NSI parameters, for a large set of pairs of parameters. We found that the best choices are the ones that combine nuclei with a different proton to neutron ratio. In particular, we identified $\rm CsI+Si$ (together with $\rm CsI + C_3F_8$, $\rm Xe + C_3F_8$ and $\rm Xe + Si$) as one of the most efficient detector combinations, able to reduce the degeneracies between certain pairs of NSI parameters to a small region in the corresponding parameter space.
Finally, we note that a longer running time than the one considered in this paper (3 years) would further improve the sensitivity of these combinations of detectors to NSI parameters, either taken individually or two at a time.


\subsection*{ACKNOWLEDGMENTS}
\vspace*{-.3cm}
\noindent S. S. C. acknowledges financial support from the LabEx P2IO (ANR-10-LABX-0038 - Project ``BSM-Nu'') in the framework of the ``Investissements d'Avenir'' (ANR-11-IDEX-0003-01) managed by the Agence Nationale de la Recherche (ANR), France.
The work of S. S. C. and S. L. is supported in part by the European Union's Horizon 2020 research and innovation programme under the Marie Sklodowska-Curie grant agreement No. 860881-HIDDeN.
The work of O. G. M. and G. S. G. has been supported in part by CONACYT-Mexico under grant A1-S-23238. O. G. M. has been supported by SNI (Sistema Nacional de Investigadores, Mexico). This work
has been supported by the Spanish grants PID2020-113775GB-I00 (AEI/10.13039/501100011033) and CIPROM/2021/054 (Generalitat Valenciana).

\vspace{2cm}

\providecommand{\href}[2]{#2}\begingroup\raggedright\endgroup


\begin{thebibliography}{10}

\bibitem{PhysRevD.9.1389}
D.~Z. Freedman, ``Coherent effects of a weak neutral current,''
  \href{http://dx.doi.org/10.1103/PhysRevD.9.1389}{{\em Phys. Rev. D}
  {\bfseries 9} (Mar, 1974) 1389--1392}.
  \url{https://link.aps.org/doi/10.1103/PhysRevD.9.1389}.

\bibitem{COHERENT:2015mry}
{\bfseries COHERENT} Collaboration, D.~Akimov {\em et~al.}, ``{The COHERENT
  Experiment at the Spallation Neutron Source},''
  \href{http://arxiv.org/abs/1509.08702}{{\ttfamily arXiv:1509.08702
  [physics.ins-det]}}.

\bibitem{Akimov:2017ade}
{\bfseries COHERENT} Collaboration, D.~Akimov {\em et~al.}, ``{Observation of
  Coherent Elastic Neutrino-Nucleus Scattering},''
  \href{http://dx.doi.org/10.1126/science.aao0990}{{\em Science} {\bfseries
  357} no.~6356, (2017) 1123--1126},
  \href{http://arxiv.org/abs/1708.01294}{{\ttfamily arXiv:1708.01294
  [nucl-ex]}}.

\bibitem{COHERENT:2020iec}
{\bfseries COHERENT} Collaboration, D.~Akimov {\em et~al.}, ``{First
  Measurement of Coherent Elastic Neutrino-Nucleus Scattering on Argon},''
  \href{http://dx.doi.org/10.1103/PhysRevLett.126.012002}{{\em Phys. Rev.
  Lett.} {\bfseries 126} no.~1, (2021) 012002},
  \href{http://arxiv.org/abs/2003.10630}{{\ttfamily arXiv:2003.10630
  [nucl-ex]}}.

\bibitem{Akimov:2021dab}
D.~Akimov {\em et~al.}, ``{Measurement of the Coherent Elastic Neutrino-Nucleus
  Scattering Cross Section on CsI by COHERENT},''
  \href{http://arxiv.org/abs/2110.07730}{{\ttfamily arXiv:2110.07730
  [hep-ex]}}.

\bibitem{Papoulias:2019txv}
D.~K. Papoulias, ``{COHERENT constraints after the COHERENT-2020 quenching
  factor measurement},''
  \href{http://dx.doi.org/10.1103/PhysRevD.102.113004}{{\em Phys. Rev. D}
  {\bfseries 102} no.~11, (2020) 113004},
  \href{http://arxiv.org/abs/1907.11644}{{\ttfamily arXiv:1907.11644
  [hep-ph]}}.

\bibitem{Khan:2019cvi}
A.~N. Khan and W.~Rodejohann, ``{New physics from COHERENT data with an
  improved quenching factor},''
  \href{http://dx.doi.org/10.1103/PhysRevD.100.113003}{{\em Phys. Rev. D}
  {\bfseries 100} no.~11, (2019) 113003},
  \href{http://arxiv.org/abs/1907.12444}{{\ttfamily arXiv:1907.12444
  [hep-ph]}}.

\bibitem{Cadeddu:2019eta}
M.~Cadeddu, F.~Dordei, C.~Giunti, Y.~F. Li, and Y.~Y. Zhang, ``{Neutrino,
  electroweak, and nuclear physics from COHERENT elastic neutrino-nucleus
  scattering with refined quenching factor},''
  \href{http://dx.doi.org/10.1103/PhysRevD.101.033004}{{\em Phys. Rev. D}
  {\bfseries 101} no.~3, (2020) 033004},
  \href{http://arxiv.org/abs/1908.06045}{{\ttfamily arXiv:1908.06045
  [hep-ph]}}.

\bibitem{Miranda:2020tif}
O.~Miranda, D.~Papoulias, G.~Sanchez~Garcia, O.~Sanders, M.~Tórtola, and
  J.~Valle, ``{Implications of the first detection of coherent elastic
  neutrino-nucleus scattering (CEvNS) with Liquid Argon},''
  \href{http://dx.doi.org/10.1007/JHEP05(2020)130}{{\em JHEP} {\bfseries 05}
  (2020) 130}, \href{http://arxiv.org/abs/2003.12050}{{\ttfamily
  arXiv:2003.12050 [hep-ph]}}.

\bibitem{Cadeddu:2021ijh}
M.~Cadeddu, N.~Cargioli, F.~Dordei, C.~Giunti, Y.~F. Li, E.~Picciau, C.~A.
  Ternes, and Y.~Y. Zhang, ``{New insights into nuclear physics and weak mixing
  angle using electroweak probes},''
  \href{http://dx.doi.org/10.1103/PhysRevC.104.065502}{{\em Phys. Rev. C}
  {\bfseries 104} no.~6, (2021) 065502},
  \href{http://arxiv.org/abs/2102.06153}{{\ttfamily arXiv:2102.06153
  [hep-ph]}}.

\bibitem{Cadeddu:2017etk}
M.~Cadeddu, C.~Giunti, Y.~Li, and Y.~Zhang, ``{Average CsI neutron density
  distribution from COHERENT data},''
  \href{http://dx.doi.org/10.1103/PhysRevLett.120.072501}{{\em Phys. Rev.
  Lett.} {\bfseries 120} no.~7, (2018) 072501},
  \href{http://arxiv.org/abs/1710.02730}{{\ttfamily arXiv:1710.02730
  [hep-ph]}}.

\bibitem{Canas:2019fjw}
B.~C. Canas, E.~A. Garces, O.~G. Miranda, A.~Parada, and G.~Sanchez~Garcia,
  ``{Interplay between nonstandard and nuclear constraints in coherent elastic
  neutrino-nucleus scattering experiments},''
  \href{http://dx.doi.org/10.1103/PhysRevD.101.035012}{{\em Phys. Rev. D}
  {\bfseries 101} no.~3, (2020) 035012},
  \href{http://arxiv.org/abs/1911.09831}{{\ttfamily arXiv:1911.09831
  [hep-ph]}}.

\bibitem{Coloma:2020nhf}
P.~Coloma, I.~Esteban, M.~C. Gonzalez-Garcia, and J.~Menendez, ``{Determining
  the nuclear neutron distribution from Coherent Elastic neutrino-Nucleus
  Scattering: current results and future prospects},''
  \href{http://dx.doi.org/10.1007/JHEP08(2020)030}{{\em JHEP} {\bfseries 08}
  no.~08, (2020) 030}, \href{http://arxiv.org/abs/2006.08624}{{\ttfamily
  arXiv:2006.08624 [hep-ph]}}.

\bibitem{Barranco:2005yy}
J.~Barranco, O.~G. Miranda, and T.~I. Rashba, ``{Probing new physics with
  coherent neutrino scattering off nuclei},''
  \href{http://dx.doi.org/10.1088/1126-6708/2005/12/021}{{\em JHEP} {\bfseries
  12} (2005) 021}, \href{http://arxiv.org/abs/hep-ph/0508299}{{\ttfamily
  arXiv:hep-ph/0508299}}.

\bibitem{Scholberg:2005qs}
K.~Scholberg, ``{Prospects for measuring coherent neutrino-nucleus elastic
  scattering at a stopped-pion neutrino source},''
  \href{http://dx.doi.org/10.1103/PhysRevD.73.033005}{{\em Phys. Rev. D}
  {\bfseries 73} (2006) 033005},
  \href{http://arxiv.org/abs/hep-ex/0511042}{{\ttfamily arXiv:hep-ex/0511042}}.

\bibitem{Liao:2017uzy}
J.~Liao and D.~Marfatia, ``{COHERENT constraints on nonstandard neutrino
  interactions},'' \href{http://dx.doi.org/10.1016/j.physletb.2017.10.046}{{\em
  Phys. Lett. B} {\bfseries 775} (2017) 54--57},
  \href{http://arxiv.org/abs/1708.04255}{{\ttfamily arXiv:1708.04255
  [hep-ph]}}.

\bibitem{Giunti:2019xpr}
C.~Giunti, ``{General COHERENT constraints on neutrino nonstandard
  interactions},'' \href{http://dx.doi.org/10.1103/PhysRevD.101.035039}{{\em
  Phys. Rev. D} {\bfseries 101} no.~3, (2020) 035039},
  \href{http://arxiv.org/abs/1909.00466}{{\ttfamily arXiv:1909.00466
  [hep-ph]}}.

\bibitem{Denton:2020hop}
P.~B. Denton and J.~Gehrlein, ``{A Statistical Analysis of the COHERENT Data
  and Applications to New Physics},''
  \href{http://dx.doi.org/10.1007/JHEP04(2021)266}{{\em JHEP} {\bfseries 04}
  (2021) 266}, \href{http://arxiv.org/abs/2008.06062}{{\ttfamily
  arXiv:2008.06062 [hep-ph]}}.

\bibitem{Galindo-Uribarri:2020huw}
A.~Galindo-Uribarri, O.~G. Miranda, and G.~S. Garcia, ``{Novel approach for the
  study of coherent elastic neutrino-nucleus scattering},''
  \href{http://dx.doi.org/10.1103/PhysRevD.105.033001}{{\em Phys. Rev. D}
  {\bfseries 105} no.~3, (2022) 033001},
  \href{http://arxiv.org/abs/2011.10230}{{\ttfamily arXiv:2011.10230
  [hep-ph]}}.

\bibitem{Khan:2021wzy}
A.~N. Khan, D.~W. McKay, and W.~Rodejohann, ``{CP-violating and charged current
  neutrino nonstandard interactions in CE\ensuremath{\nu}NS},''
  \href{http://dx.doi.org/10.1103/PhysRevD.104.015019}{{\em Phys. Rev. D}
  {\bfseries 104} no.~1, (2021) 015019},
  \href{http://arxiv.org/abs/2104.00425}{{\ttfamily arXiv:2104.00425
  [hep-ph]}}.

\bibitem{Dutta:2015nlo}
B.~Dutta, Y.~Gao, R.~Mahapatra, N.~Mirabolfathi, L.~E. Strigari, and J.~W.
  Walker, ``{Sensitivity to oscillation with a sterile fourth generation
  neutrino from ultra-low threshold neutrino-nucleus coherent scattering},''
  \href{http://dx.doi.org/10.1103/PhysRevD.94.093002}{{\em Phys. Rev. D}
  {\bfseries 94} no.~9, (2016) 093002},
  \href{http://arxiv.org/abs/1511.02834}{{\ttfamily arXiv:1511.02834
  [hep-ph]}}.

\bibitem{Lindner:2016wff}
M.~Lindner, W.~Rodejohann, and X.-J. Xu, ``{Coherent Neutrino-Nucleus
  Scattering and new Neutrino Interactions},''
  \href{http://dx.doi.org/10.1007/JHEP03(2017)097}{{\em JHEP} {\bfseries 03}
  (2017) 097}, \href{http://arxiv.org/abs/1612.04150}{{\ttfamily
  arXiv:1612.04150 [hep-ph]}}.

\bibitem{AristizabalSierra:2018eqm}
D.~Aristizabal~Sierra, V.~De~Romeri, and N.~Rojas, ``{COHERENT analysis of
  neutrino generalized interactions},''
  \href{http://dx.doi.org/10.1103/PhysRevD.98.075018}{{\em Phys. Rev. D}
  {\bfseries 98} (2018) 075018},
  \href{http://arxiv.org/abs/1806.07424}{{\ttfamily arXiv:1806.07424
  [hep-ph]}}.

\bibitem{Flores:2021kzl}
L.~J. Flores, N.~Nath, and E.~Peinado, ``{CE\ensuremath{\nu}NS as a probe of
  flavored generalized neutrino interactions},''
  \href{http://dx.doi.org/10.1103/PhysRevD.105.055010}{{\em Phys. Rev. D}
  {\bfseries 105} no.~5, (2022) 055010},
  \href{http://arxiv.org/abs/2112.05103}{{\ttfamily arXiv:2112.05103
  [hep-ph]}}.

\bibitem{Farzan:2018gtr}
Y.~Farzan, M.~Lindner, W.~Rodejohann, and X.-J. Xu, ``{Probing neutrino
  coupling to a light scalar with coherent neutrino scattering},''
  \href{http://dx.doi.org/10.1007/JHEP05(2018)066}{{\em JHEP} {\bfseries 05}
  (2018) 066}, \href{http://arxiv.org/abs/1802.05171}{{\ttfamily
  arXiv:1802.05171 [hep-ph]}}.

\bibitem{Denton:2018xmq}
P.~B. Denton, Y.~Farzan, and I.~M. Shoemaker, ``{Testing large non-standard
  neutrino interactions with arbitrary mediator mass after COHERENT data},''
  \href{http://dx.doi.org/10.1007/JHEP07(2018)037}{{\em JHEP} {\bfseries 07}
  (2018) 037}, \href{http://arxiv.org/abs/1804.03660}{{\ttfamily
  arXiv:1804.03660 [hep-ph]}}.

\bibitem{Flores:2020lji}
L.~J. Flores, N.~Nath, and E.~Peinado, ``{Non-standard neutrino interactions in
  U(1)' model after COHERENT data},''
  \href{http://dx.doi.org/10.1007/JHEP06(2020)045}{{\em JHEP} {\bfseries 06}
  (2020) 045}, \href{http://arxiv.org/abs/2002.12342}{{\ttfamily
  arXiv:2002.12342 [hep-ph]}}.

\bibitem{Cadeddu:2020nbr}
M.~Cadeddu, N.~Cargioli, F.~Dordei, C.~Giunti, Y.~F. Li, E.~Picciau, and Y.~Y.
  Zhang, ``{Constraints on light vector mediators through coherent elastic
  neutrino nucleus scattering data from COHERENT},''
  \href{http://dx.doi.org/10.1007/JHEP01(2021)116}{{\em JHEP} {\bfseries 01}
  (2021) 116}, \href{http://arxiv.org/abs/2008.05022}{{\ttfamily
  arXiv:2008.05022 [hep-ph]}}.

\bibitem{Banerjee:2021laz}
H.~Banerjee, B.~Dutta, and S.~Roy, ``{Probing
  L\ensuremath{\mu}-L\ensuremath{\tau} models with CE\ensuremath{\nu}NS: A new
  look at the combined COHERENT CsI and Ar data},''
  \href{http://dx.doi.org/10.1103/PhysRevD.104.015015}{{\em Phys. Rev. D}
  {\bfseries 104} no.~1, (2021) 015015},
  \href{http://arxiv.org/abs/2103.10196}{{\ttfamily arXiv:2103.10196
  [hep-ph]}}.

\bibitem{delaVega:2021wpx}
L.~M.~G. de~la Vega, L.~J. Flores, N.~Nath, and E.~Peinado, ``{Complementarity
  between dark matter direct searches and CE\ensuremath{\nu}NS experiments in
  U(1)' models},'' \href{http://dx.doi.org/10.1007/JHEP09(2021)146}{{\em JHEP}
  {\bfseries 09} (2021) 146}, \href{http://arxiv.org/abs/2107.04037}{{\ttfamily
  arXiv:2107.04037 [hep-ph]}}.

\bibitem{Bertuzzo:2021opb}
E.~Bertuzzo, G.~Grilli~di Cortona, and L.~M.~D. Ramos, ``{Probing light vector
  mediators with coherent scattering at future facilities},''
  \href{http://dx.doi.org/10.1007/JHEP06(2022)075}{{\em JHEP} {\bfseries 06}
  (2022) 075}, \href{http://arxiv.org/abs/2112.04020}{{\ttfamily
  arXiv:2112.04020 [hep-ph]}}.

\bibitem{Billard:2018jnl}
J.~Billard, J.~Johnston, and B.~J. Kavanagh, ``{Prospects for exploring New
  Physics in Coherent Elastic Neutrino-Nucleus Scattering},''
  \href{http://dx.doi.org/10.1088/1475-7516/2018/11/016}{{\em JCAP} {\bfseries
  11} (2018) 016}, \href{http://arxiv.org/abs/1805.01798}{{\ttfamily
  arXiv:1805.01798 [hep-ph]}}.

\bibitem{Arcadi:2019uif}
G.~Arcadi, M.~Lindner, J.~Martins, and F.~S. Queiroz, ``{New physics probes:
  Atomic parity violation, polarized electron scattering and neutrino-nucleus
  coherent scattering},''
  \href{http://dx.doi.org/10.1016/j.nuclphysb.2020.115158}{{\em Nucl. Phys. B}
  {\bfseries 959} (2020) 115158},
  \href{http://arxiv.org/abs/1906.04755}{{\ttfamily arXiv:1906.04755
  [hep-ph]}}.

\bibitem{Kosmas:2017zbh}
T.~S. Kosmas, D.~K. Papoulias, M.~Tortola, and J.~W.~F. Valle, ``{Probing light
  sterile neutrino signatures at reactor and Spallation Neutron Source neutrino
  experiments},'' \href{http://dx.doi.org/10.1103/PhysRevD.96.063013}{{\em
  Phys. Rev. D} {\bfseries 96} no.~6, (2017) 063013},
  \href{http://arxiv.org/abs/1703.00054}{{\ttfamily arXiv:1703.00054
  [hep-ph]}}.

\bibitem{Miranda:2019skf}
O.~G. Miranda, G.~Sanchez~Garcia, and O.~Sanders, ``{Coherent elastic
  neutrino-nucleus scattering as a precision test for the Standard Model and
  beyond: the COHERENT proposal case},''
  \href{http://dx.doi.org/10.1155/2019/3902819}{{\em Adv. High Energy Phys.}
  {\bfseries 2019} (2019) 3902819},
  \href{http://arxiv.org/abs/1902.09036}{{\ttfamily arXiv:1902.09036
  [hep-ph]}}.

\bibitem{Miranda:2020syh}
O.~G. Miranda, D.~K. Papoulias, O.~Sanders, M.~T\'ortola, and J.~W.~F. Valle,
  ``{Future CEvNS experiments as probes of lepton unitarity and light-sterile
  neutrinos},'' \href{http://dx.doi.org/10.1103/PhysRevD.102.113014}{{\em Phys.
  Rev. D} {\bfseries 102} (2020) 113014},
  \href{http://arxiv.org/abs/2008.02759}{{\ttfamily arXiv:2008.02759
  [hep-ph]}}.

\bibitem{Khan:2022bcl}
A.~N. Khan, ``{Extra dimensions with light and heavy neutral leptons: an
  application to CE\ensuremath{\nu}NS},''
  \href{http://dx.doi.org/10.1007/JHEP01(2023)052}{{\em JHEP} {\bfseries 01}
  (2023) 052}, \href{http://arxiv.org/abs/2208.09584}{{\ttfamily
  arXiv:2208.09584 [hep-ph]}}.

\bibitem{Cadeddu:2020lky}
M.~Cadeddu, F.~Dordei, C.~Giunti, Y.~F. Li, E.~Picciau, and Y.~Y. Zhang,
  ``{Physics results from the first COHERENT observation of coherent elastic
  neutrino-nucleus scattering in argon and their combination with cesium-iodide
  data},'' \href{http://dx.doi.org/10.1103/PhysRevD.102.015030}{{\em Phys. Rev.
  D} {\bfseries 102} no.~1, (2020) 015030},
  \href{http://arxiv.org/abs/2005.01645}{{\ttfamily arXiv:2005.01645
  [hep-ph]}}.

\bibitem{Miranda:2019wdy}
O.~G. Miranda, D.~K. Papoulias, M.~T\'ortola, and J.~W.~F. Valle, ``{Probing
  neutrino transition magnetic moments with coherent elastic neutrino-nucleus
  scattering},'' \href{http://dx.doi.org/10.1007/JHEP07(2019)103}{{\em JHEP}
  {\bfseries 07} (2019) 103}, \href{http://arxiv.org/abs/1905.03750}{{\ttfamily
  arXiv:1905.03750 [hep-ph]}}.

\bibitem{AristizabalSierra:2021kht}
D.~Aristizabal~Sierra, V.~De~Romeri, L.~J. Flores, and D.~K. Papoulias,
  ``{Impact of COHERENT measurements, cross section uncertainties and new
  interactions on the neutrino floor},''
  \href{http://dx.doi.org/10.1088/1475-7516/2022/01/055}{{\em JCAP} {\bfseries
  01} no.~01, (2022) 055}, \href{http://arxiv.org/abs/2109.03247}{{\ttfamily
  arXiv:2109.03247 [hep-ph]}}.

\bibitem{vonRaesfeld:2021gxl}
C.~von Raesfeld and P.~Huber, ``{Use of CEvNS to monitor spent nuclear fuel},''
  \href{http://dx.doi.org/10.1103/PhysRevD.105.056002}{{\em Phys. Rev. D}
  {\bfseries 105} no.~5, (2022) 056002},
  \href{http://arxiv.org/abs/2111.15398}{{\ttfamily arXiv:2111.15398
  [hep-ph]}}.

\bibitem{CCM:2021leg}
{\bfseries CCM} Collaboration, A.~A. Aguilar-Arevalo {\em et~al.}, ``{First
  Dark Matter Search Results From Coherent CAPTAIN-Mills},''
  \href{http://arxiv.org/abs/2105.14020}{{\ttfamily arXiv:2105.14020
  [hep-ex]}}.

\bibitem{Baxter:2019mcx}
D.~Baxter {\em et~al.}, ``{Coherent Elastic Neutrino-Nucleus Scattering at the
  European Spallation Source},''
  \href{http://dx.doi.org/10.1007/JHEP02(2020)123}{{\em JHEP} {\bfseries 02}
  (2020) 123}, \href{http://arxiv.org/abs/1911.00762}{{\ttfamily
  arXiv:1911.00762 [physics.ins-det]}}.

\bibitem{CONUS:2021dwh}
{\bfseries CONUS} Collaboration, H.~Bonet {\em et~al.}, ``{Novel constraints on
  neutrino physics beyond the standard model from the CONUS experiment},''
  \href{http://dx.doi.org/10.1007/JHEP05(2022)085}{{\em JHEP} {\bfseries 05}
  (2022) 085}, \href{http://arxiv.org/abs/2110.02174}{{\ttfamily
  arXiv:2110.02174 [hep-ph]}}.

\bibitem{CONNIE:2016nav}
{\bfseries CONNIE} Collaboration, A.~Aguilar-Arevalo {\em et~al.}, ``{Results
  of the Engineering Run of the Coherent Neutrino Nucleus Interaction
  Experiment (CONNIE)},''
  \href{http://dx.doi.org/10.1088/1748-0221/11/07/P07024}{{\em JINST}
  {\bfseries 11} no.~07, (2016) P07024},
  \href{http://arxiv.org/abs/1604.01343}{{\ttfamily arXiv:1604.01343
  [physics.ins-det]}}.

\bibitem{NUCLEUS:2019igx}
{\bfseries NUCLEUS} Collaboration, G.~Angloher {\em et~al.}, ``{Exploring
  $\hbox {CE}\nu \hbox {NS}$ with NUCLEUS at the Chooz nuclear power plant},''
  \href{http://dx.doi.org/10.1140/epjc/s10052-019-7454-4}{{\em Eur. Phys. J. C}
  {\bfseries 79} no.~12, (2019) 1018},
  \href{http://arxiv.org/abs/1905.10258}{{\ttfamily arXiv:1905.10258
  [physics.ins-det]}}.

\bibitem{Billard:2016giu}
J.~Billard {\em et~al.}, ``{Coherent Neutrino Scattering with Low Temperature
  Bolometers at Chooz Reactor Complex},''
  \href{http://dx.doi.org/10.1088/1361-6471/aa83d0}{{\em J. Phys. G} {\bfseries
  44} no.~10, (2017) 105101}, \href{http://arxiv.org/abs/1612.09035}{{\ttfamily
  arXiv:1612.09035 [physics.ins-det]}}.

\bibitem{Tomalak:2020zfh}
O.~Tomalak, P.~Machado, V.~Pandey, and R.~Plestid, ``{Flavor-dependent
  radiative corrections in coherent elastic neutrino-nucleus scattering},''
  \href{http://dx.doi.org/10.1007/JHEP02(2021)097}{{\em JHEP} {\bfseries 02}
  (2021) 097}, \href{http://arxiv.org/abs/2011.05960}{{\ttfamily
  arXiv:2011.05960 [hep-ph]}}.

\bibitem{Erler:2017knj}
J.~Erler and R.~Ferro-Hern\'andez, ``{Weak Mixing Angle in the Thomson
  Limit},'' \href{http://dx.doi.org/10.1007/JHEP03(2018)196}{{\em JHEP}
  {\bfseries 03} (2018) 196}, \href{http://arxiv.org/abs/1712.09146}{{\ttfamily
  arXiv:1712.09146 [hep-ph]}}.

\bibitem{PhysRevD.98.030001}
{\bfseries Particle Data Group} Collaboration, ``Review of particle physics,''
  \href{http://dx.doi.org/10.1103/PhysRevD.98.030001}{{\em Phys. Rev. D}
  {\bfseries 98} (Aug, 2018) 030001}.
  \url{https://link.aps.org/doi/10.1103/PhysRevD.98.030001}.

\bibitem{Helm:1956zz}
R.~H. Helm, ``{Inelastic and Elastic Scattering of 187-Mev Electrons from
  Selected Even-Even Nuclei},''
  \href{http://dx.doi.org/10.1103/PhysRev.104.1466}{{\em Phys. Rev.} {\bfseries
  104} (1956) 1466--1475}.

\bibitem{Papoulias:2018uzy}
D.~K. Papoulias, R.~Sahu, T.~S. Kosmas, V.~K.~B. Kota, and B.~Nayak, ``{Novel
  neutrino-floor and dark matter searches with deformed shell model
  calculations},'' \href{http://dx.doi.org/10.1155/2018/6031362}{{\em Adv. High
  Energy Phys.} {\bfseries 2018} (2018) 6031362},
  \href{http://arxiv.org/abs/1804.11319}{{\ttfamily arXiv:1804.11319
  [hep-ph]}}.

\bibitem{Papoulias:2019lfi}
D.~K. Papoulias, T.~S. Kosmas, R.~Sahu, V.~K.~B. Kota, and M.~Hota,
  ``{Constraining nuclear physics parameters with current and future COHERENT
  data},'' \href{http://dx.doi.org/10.1016/j.physletb.2019.135133}{{\em Phys.
  Lett. B} {\bfseries 800} (2020) 135133},
  \href{http://arxiv.org/abs/1903.03722}{{\ttfamily arXiv:1903.03722
  [hep-ph]}}.

\bibitem{Hoferichter:2020osn}
M.~Hoferichter, J.~Men\'endez, and A.~Schwenk, ``{Coherent elastic
  neutrino-nucleus scattering: EFT analysis and nuclear responses},''
  \href{http://dx.doi.org/10.1103/PhysRevD.102.074018}{{\em Phys. Rev. D}
  {\bfseries 102} no.~7, (2020) 074018},
  \href{http://arxiv.org/abs/2007.08529}{{\ttfamily arXiv:2007.08529
  [hep-ph]}}.

\bibitem{Friedrich:1982esq}
J.~Friedrich and N.~Voegler, ``{The salient features of charge density
  distributions of medium and heavy even-even nuclei determined from a
  systematic analysis of elastic electron scattering form factors},''
  \href{http://dx.doi.org/10.1016/0375-9474(82)90147-6}{{\em Nucl. Phys. A}
  {\bfseries 373} (1982) 192--224}.

\bibitem{Wolfenstein:1977ue}
L.~Wolfenstein, ``{Neutrino Oscillations in Matter},''
\href{http://dx.doi.org/10.1103/PhysRevD.17.2369}{{\em Phys.Rev.} {\bfseries
  D17} (1978) 2369--2374}.

\bibitem{Angeli:2013epw}
I.~Angeli and K.~P. Marinova, ``{Table of experimental nuclear ground state
  charge radii: An update},''
  \href{http://dx.doi.org/10.1016/j.adt.2011.12.006}{{\em Atom. Data Nucl. Data
  Tabl.} {\bfseries 99} no.~1, (2013) 69--95}.

\bibitem{Mikheyev:1985zog}
S.~P. Mikheyev and A.~Y. Smirnov, ``{Resonance Amplification of Oscillations in
  Matter and Spectroscopy of Solar Neutrinos},'' {\em Sov. J. Nucl. Phys.}
  {\bfseries 42} (1985) 913--917.

\bibitem{Miranda:2004nb}
O.~G. Miranda, M.~A. Tortola, and J.~W.~F. Valle, ``{Are solar neutrino
  oscillations robust?},''
  \href{http://dx.doi.org/10.1088/1126-6708/2006/10/008}{{\em JHEP} {\bfseries
  10} (2006) 008}, \href{http://arxiv.org/abs/hep-ph/0406280}{{\ttfamily
  arXiv:hep-ph/0406280}}.

\bibitem{Coloma:2016gei}
P.~Coloma and T.~Schwetz, ``{Generalized mass ordering degeneracy in neutrino
  oscillation experiments},''
  \href{http://dx.doi.org/10.1103/PhysRevD.94.055005}{{\em Phys. Rev. D}
  {\bfseries 94} no.~5, (2016) 055005},
  \href{http://arxiv.org/abs/1604.05772}{{\ttfamily arXiv:1604.05772
  [hep-ph]}}. [Erratum: Phys.Rev.D 95, 079903 (2017)].

\bibitem{Esteban:2018ppq}
I.~Esteban, M.~C. Gonzalez-Garcia, M.~Maltoni, I.~Martinez-Soler, and
  J.~Salvado, ``{Updated constraints on non-standard interactions from global
  analysis of oscillation data},''
  \href{http://dx.doi.org/10.1007/JHEP08(2018)180}{{\em JHEP} {\bfseries 08}
  (2018) 180}, \href{http://arxiv.org/abs/1805.04530}{{\ttfamily
  arXiv:1805.04530 [hep-ph]}}. [Addendum: JHEP 12, 152 (2020)].

\bibitem{Coloma:2019mbs}
P.~Coloma, I.~Esteban, M.~C. Gonzalez-Garcia, and M.~Maltoni, ``{Improved
  global fit to Non-Standard neutrino Interactions using COHERENT energy and
  timing data},'' \href{http://dx.doi.org/10.1007/JHEP02(2020)023}{{\em JHEP}
  {\bfseries 02} (2020) 023}, \href{http://arxiv.org/abs/1911.09109}{{\ttfamily
  arXiv:1911.09109 [hep-ph]}}. [Addendum: JHEP 12, 071 (2020)].

\bibitem{Chaves:2021pey}
M.~Chaves and T.~Schwetz, ``{Resolving the LMA-dark NSI degeneracy with
  coherent neutrino-nucleus scattering},''
  \href{http://dx.doi.org/10.1007/JHEP05(2021)042}{{\em JHEP} {\bfseries 05}
  (2021) 042}, \href{http://arxiv.org/abs/2102.11981}{{\ttfamily
  arXiv:2102.11981 [hep-ph]}}.

\end{thebibliography}
\end{document}